\documentclass[11pt]{article}
\usepackage{amsmath}

\usepackage{amssymb,color}
\usepackage{epsfig}

\newcommand{\N}{N\raise.7ex\hbox{\underline{$\circ $}}$\;$}

\renewcommand{\theequation}{\thesection.\theequation}
\numberwithin{equation}{section}

\begin{document}

\title{On geometry influence  on the behavior of a quantum mechanical
scalar  particle with  intrinsic structure in external magnetic
and electric fields
 }

\maketitle

\author{
O.V. Veko\footnote{Kalinkovichi Gymnasium,
Belarus,vekoolga@mail.ru}, K.V.  Kazmerchuk\footnote{Mosyr State
Pedagogical University, Belarus, kristinash2@mail.ru},
 E.M. Ovsiyuk\footnote{Mosyr State Pedagogical University, Belarus, e.ovsiyuk@mail.ru},
V.V. Kisel\footnote{Belarusian State University of Informatics and
Radioelectronics},
 V.M. Red'kov\footnote{B.I. Stepanov Institute of Physics, NAS of Belarus, redkov@dragon.bas-net.by}}

\begin{abstract}

Relativistic theory of the Cox's  scalar  not point-like particle
with intrinsic  structure is developed on the background of
arbitrary curved space-time. It is shown that in the most general
form,  the  extended  Proca-like tensor first order
 system of equations contains non minimal interaction terms through electromagnetic tensor $F_{\alpha \beta}$
 and Ricci tensor $R_{\alpha \beta}$.

 In
relativistic Cox's theory,  the limiting procedure to
non-relativistic approximation is performed in a special class of
curved space-time models.
   This theory is specified in simple geometrical backgrounds:
   Euclid's, Lobache\-vsky's, and Rie\-mann's.
  Wave equation  for the Cox's particle
is solved  exactly in presence of external uniform magnetic and
electric fields in the case of Minkowski space. Non-trivial
additional  structure of the particle modifies the frequency of a
quantum oscillator arising effectively in presence if external
magnetic field. Extension of these problems to the case of the
hyperbolic Lobachevsky  space is examined.
 In
presence of  the magnetic field, the quantum problem in radial
variable has been solved exactly; the quantum motion in
z-direction is described by 1-dimensional Schr\"{o}dinger-like
equation in an effective potential which turns out to be  too
difficult for analytical treatment. In the presence of electric
field, the situation is similar. The same analysis has been
performed for spherical Riemann space model.

\end{abstract}




\section{ Scalar Cox's particle with intrinsic structure}

In 1982 W. Cox \cite{Cox-1982}
 proposed a special  wave equation for a scalar particle
with a larger set of tensor components  than  the usual
Proca's approach includes:  he  used the set of a  scalar, 4-vector,
antisymmetric and (irreducible) symmetric tensor,  thus starting
with the 20-component wave function (see Section {\bf 13}).

First, let us consider the system of Cox's equations
\cite{Cox-1982}   in the Minkowski
space.
 We  will use a Proca-like generalized system    obtained after elimination from
  the initial system of Cox's equations  two second-rank tensors (see Section {\bf 13}):
\begin{eqnarray}
 \left ({i \over \sqrt{-g} } \; { \partial \over  \partial x^{\alpha} }
\sqrt{-g}  - {e \over c \hbar }  \; A_{\alpha} \right )\; g^{\alpha
\beta} \Phi_{\beta} = {Mc \over  \hbar }    \Phi
\; ,
\quad
 K_{\rho}^{\;\;\alpha} \left (i\;  \partial _{\alpha } -  {e\over c \hbar} A_{\alpha} \right )
\Phi = {Mc \over  \hbar }    \Phi_{\rho} \; .\quad
\label{start}
\end{eqnarray}

\noindent
$K_{\rho}^{\;\;\alpha}$ is a tensor inverse to
$\Lambda_{\sigma}^{\;\;\alpha}= M c
\delta_{\sigma}^{\;\;\alpha}+\lambda F_{\sigma}^{\;\;\alpha}$ ($\lambda$ stands for additional parameter
responsible for non-trivia intrinsic structure of a scalar particle in Cox's approach):

 \begin{eqnarray}
K_{\rho}^{\;\;\alpha}= \lambda_{1}\; \delta_{\alpha}^{\;\; \beta}
+ \lambda_{2} \; F_{\alpha}^{\;\;\beta}
+ \lambda_{3} \;
F_{\alpha}^{\;\; \rho}F_{\rho}^{\;\; \beta}+ \lambda_{4} \;
F_{\alpha}^{\;\;\rho} F_{\rho}^{\;\; \sigma} F_{\sigma}^{\;\;
\beta} \; ,
\end{eqnarray}

\noindent  $\lambda_{i}$ are  expressed through electromagnetic
invariants (for more technical details see Section {\bf 13}).

In geometrical models with metrics of  special  type
$
dS^{2} = c^{2} dt^{2} + g_{kl}(x)\;  dx^{k} dx^{l} \; ,
$
one cap perform  non-relativistic approximation and derive
extended Schr\"{o}dinger type  equation (see Section {\bf 13}):
 \begin{eqnarray}
 D_{t}    \;  \Psi ={1 \over 2M}
 \stackrel{\circ}{D}_{k}     (-g^{kj})
\left ( K_{j}^{\;\;l} D_{l}  + Mc  K_{j}^{\;\;0} \right ) \Psi
-
 {1 \over 2} \left (   (K_{0}^{\;\;0} -1)  Mc^{2}   +
 K_{0}^{\;\;j} c D_{j}   \right ) \Psi \; ,\quad
\label{1}
\end{eqnarray}

\noindent where the notation is used
\begin{eqnarray}
i \hbar   \;  \partial  _{t }    -  e  A_{0} = D_{t} \; ,\quad
 i c
\hbar   \;  \partial  _{k }    -  e  A_{k} = c  D_{k} \; ,
\quad {i c\hbar  \over \sqrt{-g}} \; {\partial \over  \partial x^{k} }
\sqrt{-g}  - e  \; A_{k} = c \;\stackrel{\circ}{D}_{k} \; ,
\nonumber
\end{eqnarray}

\noindent
It is a generalized Schr\"{o}dinger equation for the
 particle  with   intrinsic structure.

In presence of a pure magnetic field, the above equation (\ref{1}) takes a more simple form
\begin{eqnarray}
 D_{t}    \;  \Psi = - {1 \over 2M}
 \stackrel{\circ}{D}_{k}     g^{kj}(x) \stackrel{\ast}{D}_{j}    \Psi \; ,
\label{magnetic}
\end{eqnarray}

\noindent where the notation is used (let $\Gamma = \lambda /  mc
$)
\begin{eqnarray}
\stackrel{\ast}{D}_{1}= K_{1}^{\;\;l} D_{l} ={1 \over 1  +
\Gamma^{2} B_{i} B^{i} } \left  [  D_{1} + \Gamma (B_{2} D^{3} -
B_{3} D^{2}) +   \Gamma^{2} B^{1} \; (B_{i} D_{i} )  \right ]
,
\nonumber
\\
\stackrel{\ast}{D}_{2} = K_{2}^{\;\;l} D_{l} = {1 \over 1  +
\Gamma^{2} B_{i}B^{i} } \left [   D_{2} + \Gamma (B_{3} D^{1} -
B_{1} D^{3}) +   \Gamma^{2} B^{2} \; (B_{i} D_{i} ) \right  ]
 ,
\nonumber
\\
\stackrel{\ast}{D}_{3}  = K_{3}^{\;\;l} D_{l} = {1 \over 1 +
\Gamma^{2} B_{i}B^{i}} \left [
 D_{3} + \Gamma (B_{1} D^{2} - B_{2} D^{1}) +   \Gamma^{2} B^{3} \; (B_{i} D_{i} ) \right  ] ,
\nonumber
\\
g^{22}g^{33} B_{1} = B^{1} \; ,\; \;\; g^{33}g^{11} B_{2} = B^{2}
\; ,\;\;\;
 g^{11}g^{22} B_{3} = B^{3} \; ,\;\;\;
F_{ij}=\epsilon_{ijk}B_{k} \; .
\nonumber
\end{eqnarray}

In presence of a pure  electric field,
the above equation (\ref{1}) takes the  form
\begin{eqnarray}
\left (  D_{t}  - c\;
 {   \Gamma ^{2} E_{i}E^{i}  Mc  +
     \Gamma    E^{j} D_{j}  \over 2( 1 +  \Gamma^{2} E_{i}E^{i}  )}  \right )     \Psi
    =
  {1 \over 2M}
 \stackrel{\circ}{D}_{k}     (-g^{kj})
\left ( D_{j}  + { \Gamma^{2}  E_{j} (E^{i} D_{i})   +
 Mc   \Gamma E_{j}  \over 1 + \Gamma^{2} E_{i}E^{i} }   \right  ) \Psi
\end{eqnarray}

\noindent
where
$$
g^{11}E_{1} = E^{1} ,
 \qquad g^{22}E_{2} =E^{2}, \qquad  g^{33}E_{3} = E^{3}, \;\quad
E_{i}= F_{0i}\; .
$$

\section{ Cox's particle  in the magnetic field,
Minkowski  space}

Let the homogeneous  magnetic field
$
 \vec{A} = {1 \over 2}  \vec{x}
\times \vec{B} $
be  directed along the axis $z$:
\begin{eqnarray}
(A_{j})  = \vec{A} = {1 \over 2 } \left  ( x^{2} B , \;  - x^{1}
B,\;  0  \right ) \; .
\nonumber
\label{1.1a}
\end{eqnarray}

\noindent Recalculating the potential to cylindrical coordinates
by the formulas
\begin{eqnarray}
A_{j'} = { \partial x^{j} \over \partial x ^{j'} } A_{j}, \; x^{j}
= (x,y,z), \; x{j'} = (r , \phi , z )\;, \quad
x^{1} = r \cos \phi, \; x^{2} =  r \sin \phi,  \; x^{3} = x^{3'} =
z  \; , \nonumber
\end{eqnarray}

\noindent we obtain
\begin{eqnarray}
A_{r} = 0, \; A_{\phi} = - {Br^{2} \over 2}  , \; A_{z} = 0, \;
F_{r\phi} =
 -Br \; .
 \label{1.1b}
\end{eqnarray}

\noindent The metric tensor in these coordinates and field
variables are determined  by
\begin{eqnarray}
dS^{2} = c^{2} dt^{2} - dr^{2} - r^{2} d\phi^{2}  - dz^{2} , \;
\sqrt{-g} = r \; , \quad
 B_{3} =   -Br  ,\;
 B^{3} =   -Br^{-1}  ,\;  B_{i}B^{i} = B^{2}\,.\quad
\label{1.1d}
\end{eqnarray}

\noindent
The Schr\"{o}dinger equation  for this case reads
\begin{eqnarray}
 D_{t}      \Psi =  {1 \over 2M}  \left (
 \stackrel{\circ}{D}_{1}       \stackrel{\ast}{D}_{1} +
 \stackrel{\circ}{D}_{2}      r^{-2}  \stackrel{\ast}{D}_{2}+
 \stackrel{\circ}{D}_{3}       \stackrel{\ast}{D}_{3} \right  )  \Psi  ,
\nonumber
\label{1.2}
\end{eqnarray}

\noindent where
\begin{eqnarray}
D_{1} =i\hbar \partial_{r}\;  , \;  D_{2} =i\hbar
\partial_{\phi} + {e\over c } {Br^{2} \over 2} \; , \; D_{3}
=i\hbar \partial_{z} \; , \quad
\stackrel{\circ}{D}_{1} = i \hbar ( \partial_{r} + {1 \over r}
),\qquad \stackrel{\circ}{D}_{2} =i\hbar \partial_{\phi} + {e\over
c } {Br^{2} \over 2} \; , \nonumber
\\
 \stackrel{\circ}{D}_{3}
=i\hbar
\partial_{z}  \; ,\qquad
\stackrel{\ast}{D}_{1}= {1 \over 1 + \Gamma^{2} B^{2}} (D_{1}-
\Gamma B_{3}D^{2} )
 = {1 \over 1 + \Gamma^{2} B^{2}} \left (
 i \hbar \partial_{r} - {\Gamma B\over  r } (i\hbar \partial_{\phi} + {e\over c } {Br^{2} \over 2}) \right  )
 ,
\nonumber \end{eqnarray}
$$
 \stackrel{\ast}{D}_{2} =
{1 \over 1 + \Gamma^{2} B^{2}  } ( \; D_{2} +  \Gamma B_{3} D^{1}
)
=
 {1 \over 1  +  \Gamma^{2} B^{2} } \left (
(i\hbar \partial_{\phi} + {e\over c } {Br^{2} \over 2} ) + i\hbar
\Gamma B r  \partial_{r} \right )
 ,
$$
$$
\stackrel{\ast}{D}_{3} = { (  D_{3} +   \Gamma^{2} B^{3} \; B_{3}
D_{3}  ) \over 1 +  \Gamma^{2} B^{2} }  =
 i\hbar \partial_{z}
  \; ;
$$

\noindent below  we will use the notation
$$
 {eB \over 2\hbar c } = b, \qquad \Gamma B = \gamma\; .
$$

\noindent
We  compute
\begin{eqnarray}
{1 \over 2M} \stackrel{\circ}{D}_{1}       \stackrel{\ast}{D}_{1}=
 - {\hbar^{2}\over 2M(1 + \gamma^{2} )}
 \left (
    \partial_{r}^{2} + {1 \over r} \partial_{r}  - {\gamma \over r}
     \partial_{r} \partial_{\phi} +i\gamma b r \partial_{r}
      +2i\gamma b      \right )  ,
\nonumber
\\
{1 \over 2M} \stackrel{\circ}{D}_{2}       \stackrel{\ast}{D}_{2}=
 - {\hbar^{2}\over 2M(1 + \gamma^{2} )}
  \left ( {1 \over r^{2}}
(\partial_{\phi} - i b r^{2}  )^{2} +
  \gamma (\partial_{\phi} -  ib r^{2}  ) \;  {1 \over r } \partial_{r} \right )  ,
\nonumber
\\
{1 \over 2M} \stackrel{\circ}{D}_{3}
  \stackrel{\ast}{D}_{3}= - {\hbar^{2}\over 2M(1 + \gamma^{2} )}   (1 + \gamma^{2} ) \partial_{z}^{2} \; .
\label{1.4}
\end{eqnarray}

After using the substitution for the wave function
$$
\Psi = e^{-iE t / \hbar } e^{im\phi} e^{ikz} R(r)  ,
\quad \epsilon =
 {2m E   \over \hbar^{2}  }  (1 + \gamma^{2})
$$
 we get the radial Schr\"{o}dinger equation
\begin{eqnarray}
 \left [ {d^{2} \over dr^{2}}  +  {1 \over r}  {d \over dr}
 + 2i \gamma b  + \epsilon
   -     {(m -  b r^{2}  )^{2}  \over r^{2}}
    - (1 + \gamma^{2}) k^{2}  \right ] R
     = 0  .
\label{1.6}
\end{eqnarray}

\noindent By  physical reasons
 parameter $\gamma $ must be purely
imaginary (see Section {\bf 13}): $\gamma = - i \eta$; so the radial equation reads
\begin{eqnarray}
\left  [ {d^{2} \over dr^{2}}  +  {1 \over r}  {d \over dr}  + 2
\eta b  + \epsilon
   -
  {(m -  b r^{2}  )^{2}  \over r^{2}}
    - (1 - \eta^{2}) k^{2}  \right  ]  R = 0 .
\label{1.7'}
\end{eqnarray}

\noindent With the use of
notation
$
 \epsilon  - (1 - \eta^{2}) k^{2} +  2 \eta  b  =  \epsilon '\; ,
$
equation (\ref{1.7'}) can be written as
\begin{eqnarray}
\left ( {d^{2} \over dr^{2}}  +  {1 \over r}  {d \over dr}   -
  {(m -  b r^{2}  )^{2}  \over r^{2}}  +\epsilon '  \right )  = 0 \; ,
\label{1.9}
\end{eqnarray}

\noindent which  coincides with the equation arising
in the  problem of the usual particle in the
magnetic field. Its solutions are known. We present here only an
expression for the energy spectrum
\begin{eqnarray}
\epsilon = 4b ( n + { m + \mid m \mid  +1 \over 2} )
+ (1 -
\eta^{2}) k^{2} -  2\eta  b \; ; \label{1.10a}
\end{eqnarray}

\noindent from this after translating to  ordinary units we  obtain
\begin{eqnarray}
E ={p^{2} \over 2M} +  {1 \over 1 - \eta^{2} } {eB  \over M c
}\hbar \left ( n + { m + \mid m \mid  +1 \over 2} \right )
   - {\eta \over 2} {1 \over 1 - \eta^{2}}  {e B \over Mc}\hbar\;.
\label{1.10b}
\end{eqnarray}

\noindent  With the use of notation
$
\eta =  \Gamma  B , \; \Gamma^{*} = \Gamma , \; \omega =
{eB \over M c} \; ,
$
the formula for the energy levels can be written as
\begin{eqnarray}
E ={p^{2} \over 2M} +  {\omega \hbar  \over 1 - (\Gamma  B)^{2} }
 \left ( n + { m + \mid m \mid  +1 \over 2} \right )
   -  {\omega  \hbar  \over 1 - (\Gamma  B)^{2}}     {\Gamma  B \over 2} \; .
\label{1.10c}
\end{eqnarray}

\noindent Thus,  the intrinsic structure of the Cox's particle
modifies
 the frequency of the quantum oscillator (in  fact, this result was firstly produced in different formalism
 by Kisel \cite{Kisel}).
\begin{eqnarray}
\omega \;\; \Longrightarrow \;\; \tilde{\omega}  = {\omega
\over 1 - \Gamma^{2}   B^{2}} , \; \omega = {eB \over M c} \; .
\label{1.10d}
\end{eqnarray}

\section{ Cox's particle in the magnetic field in the Lobachevsky space}

In a special (cylindrical) coordinate system in the Lobachevsky
space, analogue of the uniform magnetic field is determined by the
relations   (we use dimensionless coordinate
$r$  obtained by dividing on the curvature radius $\rho$):
\begin{eqnarray}
dS^{2} = c^{2} dt^{2} - \mbox{ch}^{2} z ( dr^{2} + \mbox{sh}^{2} r
d\phi^{2} ) + dz^{2}  , \nonumber
\\
A_{\phi} = - B \rho^{2} ( \mbox{ch}\;r-1)   , \; F_{r\phi} =
  -B\rho \;\mbox{sh}\; r  \;,
\nonumber
\end{eqnarray}
\begin{eqnarray}
 B_{3} =   -B\rho\; \mbox{sh}\; r  \;,\;
 B^{3} =   -{B \over \rho\; \mbox{sh} \; r\; \mbox{ch}^{4} z} \; ,
 \quad   B_{i}B^{i} = B^{2} \mbox{ch}^{-4} z \;. \qquad
\label{2.1}
\end{eqnarray}

The wave equation in this case reads
\begin{eqnarray}
 D_{t}      \Psi =  {1 \over 2M \rho^{2}} \left (
 \stackrel{\circ}{D}_{1}       \stackrel{\ast}{D}_{1}  +
 \stackrel{\circ}{D}_{2}     {1 \over  \mbox{sh}^{2} r}  \stackrel{\ast}{D}_{2}+
 \stackrel{\circ}{D}_{3}       \stackrel{\ast}{D}_{3} \right ]\Psi  ,
 \label{2.2'}
\end{eqnarray}

\noindent where
$$
D_{1} =i \hbar   \partial_{r} ,\quad  D_{3} =i \hbar
\partial_{z}  ,\quad
 D_{2} =i\hbar
\partial_{\phi} + {e\over c } B \rho^{2} ( \mbox{ch}\;r-1) ,
$$
$$
\stackrel{\circ}{D}_{1} =
 i \hbar   ( \partial_{r} + {\mbox{ch}\; r \over \mbox{sh}\; r} ),
\quad
 \stackrel{\circ}{D}_{2}
= i\hbar \partial_{\phi} + {e\over c } B \rho^{2} (
\mbox{ch}\;r-1),\quad
 \stackrel{\circ}{D}_{3}
=i\hbar   ( \partial_{z}  + 2 { \mbox{sh}\;  z \over \mbox{ch}\;
z} )  ,
$$
$$
\stackrel{\ast}{D}_{1}=
 {1 \over 1 + \Gamma^{2} B^{2}\mbox{ch}^{-4} z  }
\left [
 i \hbar  \partial_{r}   - {\Gamma B \mbox{ch}^{-2} z \over \mbox{sh} r } (i\hbar
\partial_{\phi} + {e\over c } B \rho^{2} ( \mbox{ch}\;r-1))
  \right  ]
  ,
$$
$$
\stackrel{\ast}{D}_{2} =
 {1 \over 1  +  \Gamma^{2} B^{2} \mbox{ch}^{-4} z }
 \left [
( i\hbar \partial_{\phi}  + {e\over c } B \rho^{2} ( \mbox{ch}\;r-1) ) + i\hbar
\Gamma B  \mbox{ch}^{-2} z\; \mbox{sh}\; r  \partial_{r} \right ]
  ,
$$
$$
\stackrel{\ast}{D}_{3} = { (  D_{3}  +   \Gamma^{2} B^{3} \; B_{3}
D_{3}  )  \over 1 +  \Gamma^{2} B^{2}\mbox{ch}^{-4} z }  =
 i\hbar  \partial_{z}
 .
$$


\noindent Below the notation is used:
\begin{eqnarray}
 (eB \rho^{2} / \hbar c ) = b,  \qquad   \Gamma B  \mbox{ch}^{-2} z= \gamma (z).
 \nonumber
 \end{eqnarray}
We compute
\begin{eqnarray}
{1 \over 2M \rho ^{2}} \stackrel{\circ}{D}_{1}  g^{11}
\stackrel{\ast}{D}_{1}= - {\hbar^{2} \mbox{ch}^{-2} z \over 2 M
\rho^{2} (1 + \gamma^{2}(z) )} \nonumber
\\
   \times
    \left (      \partial_{r}^{2} + ( {\mbox{ch}\; r \over \mbox{sh}\; r } +
    i \gamma (z) b { \mbox{ch}\; r -1 \over \mbox{sh}\; r} ) \partial_{r}
   - {\gamma (z) \over \mbox{sh}\; r} \partial_{r} \partial_{\phi}  + i \gamma (z)  b     \right )  ,
\nonumber
\\
{1 \over 2M \rho^{2}} \stackrel{\circ}{D}_{2}   g^{22}
\stackrel{\ast}{D}_{2}= - {\hbar^{2} \mbox{ch}^{-2} z \over 2M
\rho^{2} (1 + \gamma^{2} (z) )} \nonumber
\\
   \times   \left [ {1 \over \mbox{sh}^{2} r }
  [ \partial_{\phi} - i b  ( \mbox{ch}\;r-1)  ]^{2}     +
  \gamma (z)  [ \partial_{\phi}  - i b  ( \mbox{ch}\;r-1)  ] \;  {1 \over \mbox{sh}\; r } \partial_{r} \right ]  ,
\nonumber
\\
{1 \over 2M\rho^{2} } \stackrel{\circ}{D}_{3}  g^{33}
\stackrel{\ast}{D}_{3}= - {\hbar^{2}\over 2M \rho^{2} }
   ( \partial_{z}  + 2 { \mbox{sh}\;  z \over \mbox{ch}\;  z} )\partial_{z}  .
\nonumber
\\
\label{2.4}
\end{eqnarray}

\noindent After using the substitution
for  the wave function:
\begin{eqnarray}
\Psi = e^{-iE t / \hbar } e^{im\phi} Z(z)  R(r) \; ,
\label{2.5}
\end{eqnarray}

\noindent   the Schr\"{o}dinger equation (\ref{2.2'}) gives (the function  $\gamma (z)$ must be  imaginary,
 $i \gamma (z\Longrightarrow \gamma (z)$)
 \begin{eqnarray}
 \left [ {\mbox{ch}^{-2} z  \over 1 - \gamma^{2}(z)} \left (
\partial^{2}_{r}  +  {\mbox{ch}\; r  \over \mbox{sh}\; r }
\partial_{r}
  - {   [ m -  b  ( \mbox{ch}\;r-1)  ]^{2} \over \mbox{sh}^{2} r }
    +  b \gamma  (z)  \right )
\right.
\nonumber
\\
\left.
     +  \epsilon +  ( \partial_{z}  + 2 { \mbox{sh}\;  z \over \mbox{ch}\;  z} )\partial_{z}
     \right ] R (r) Z(z) = 0 \; , \quad \epsilon =
 { E   \over \hbar^{2} /2 M\rho^{2}  } \; .
\label{2.7}
\end{eqnarray}

\noindent In this equation, the variables are separated:
\begin{eqnarray}
 { 1 \over R} \left ( {d^{2} \over dr^{2}}  +  {\mbox{ch}\; r  \over \mbox{sh}\; r }  {d \over dr}
  - {   [ m -  b  ( \mbox{ch}\;r-1)  ]^{2} \over \mbox{sh}^{2} r }\right )R
 \nonumber
 \\
    +
  {1 \over Z} (1 - \gamma^{2} (z)) \; \mbox{ch}^{2}z
   \left (   {b \; \gamma  (z)\; \mbox{ch}^{-2} z  \over 1 - \gamma^{2}(z)}    +
    \epsilon +  ( { d \over dz}   + 2 { \mbox{sh}\;  z \over \mbox{ch}\;  z} ) {d \over dz}
     \right )  Z = 0  .
\label{2.8}
\end{eqnarray}

\noindent The radial equation for the function $R (r)$ reads
\begin{eqnarray}
\left ( {d^{2} \over dr^{2}}  +  {\mbox{ch}\; r  \over \mbox{sh}\;
r }  {d \over dr}
  -  {   [ m -  b  ( \mbox{ch}\;r-1)  ]^{2} \over \mbox{sh}^{2} r }
    +  \Lambda  \right ) R = 0\; ;
\label{2.9'}
\end{eqnarray}

\noindent the equation for $Z(z)$ is
 (remember that
$ \gamma = B  \Gamma$)

\begin{eqnarray}
\left (
     {d^{2} \over dz^{2}}   + 2 { \mbox{sh}\;  z \over \mbox{ch}\;  z} {d \over dz}
  + \epsilon
    + {b \gamma -  \Lambda   \mbox{ch}^{2}z   \over  \mbox{ch}^{4} z  - \gamma ^{2} }  \right )  Z = 0  .
\nonumber
\label{2.10'}
\end{eqnarray}

\section{ Analysis of the equation in the variable $z$ }

In equation (\ref{2.10'}), let us  eliminate  the first derivative
term:
\begin{eqnarray}
Z ={1 \over  \mbox{ch}\; z} f(z), \quad
      U(z) = - { b \gamma - \Lambda  \; \mbox{ch}^{2}z   \over  \mbox{ch}^{4} z  - \gamma ^{2} } ,
\nonumber
\\
\left (
     {d^{2} \over dz^{2}}     + \epsilon -1  -U(z)
     \right )  f(z) = 0  .
\label{3.1'}
\end{eqnarray}

\noindent Eq. (\ref{3.1'}) can be viewed as the Schr\"{o}dinger
equation in the effective potential field $U(z)$. The
corresponding effective force is
\begin{eqnarray}
F_{z} = -{d U \over dz} \qquad\qquad \nonumber
\\
=  2 \; \mbox{ch}\; z  \; \mbox{sh}\; z \; { \Lambda \;
\mbox{ch}^{4} z - 2b\gamma \; \mbox{ch}^{2} z +\gamma^{2} \Lambda
\over (\mbox{ch}^{4} z  - \gamma ^{2})^{2} }
 . \label{3.2'}
 \end{eqnarray}

\noindent We find the points of local extremum:
 $
z=0\; $ and the roots of a quadratic equation
\begin{eqnarray}
\Lambda \; \mbox{ch}^{4} z - 2b\gamma \; \mbox{ch}^{2} z
+\gamma^{2} \Lambda =0 \qquad \Longrightarrow \nonumber
\\
\left ( \mbox{ch}^{2} z \right ) |_{1,2} = {b  \over \Lambda }
\gamma \pm
 \sqrt{({b^{2}  \over \Lambda ^{2} }   -  1) \gamma^{2}} \; .
\label{3.3'}
\end{eqnarray}

\noindent
When considering the
bound states (for motion in  the variable $r$)
 we have $\Lambda^{2}> b^{2}$. This means that the square root in (\ref{3.3'})  is an imaginary number.
  Consequently, the point of zero force (equilibrium points) except $z = 0$ cannot exist.
The situation is illustrated in the Fig. 1.

\begin{figure}[h]
\centering \includegraphics[scale=0.5]{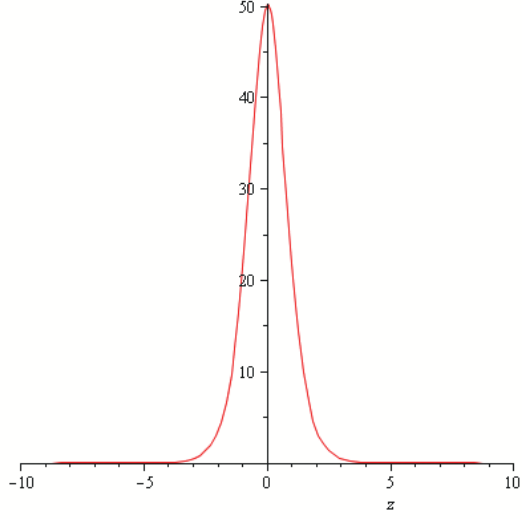} \caption{ \emph{
Effective potential $U(z)$:  } }\label{fig:1}
\end{figure}

After the change of variables $ \mbox{ch}^{2}z = y$,
 the
differential equation (\ref{3.1'}) reads
\begin{eqnarray}
\left [ {d^{2} \over dy^{2} } +\left ( {3 \over 2} {1 \over y} +{1
\over 2}{1 \over y-1 } \right ) {d \over d  y }+ { \epsilon \over
4y(y-1)}  + {b\gamma-  \Lambda  \; y   \over  (y   - \gamma) (y+
\gamma )  4 y(y-1)} \right ] Z(y) = 0 \; . \label{3.4'}
\end{eqnarray}

\noindent Note that  singular points
$y = 0, \pm \gamma (\mid \gamma \mid << 1) $
are located outside the physical range of the variable.
Further progress in analytical treatment of eq.  (\ref{3.4'}) (with
5 singular points) is hardly   possible.

\section{ Solution of the radial equation}

 Let us turn to the radial equation (\ref{2.9'}) for the
function $R (r)$.
It is solvable
in hypergeometric  functions -- see more detail \cite{Bogush-et-al}.
Below we will write done only final results on energy spectrum.
There exist only   finite series of bound states, defined by relations
\begin{eqnarray}
m < 2b  \;, \ {m +\mid m \mid \over 2} + n +1/2  \leq b \;  ,
\nonumber
\\
\Lambda -1/4 = 2b\left( {m + \mid m \mid \over 2 } + n +
1/2\right)
- \left(  { m + \mid m \mid \over 2} + n + {1\over 2}
\right)^{2} \; ;
\label{spectrum}
\end{eqnarray}

\noindent
$\Lambda$ obeys the restriction
$
b \leq \Lambda  \;.
$
 In usual units the last relation can be written as:
\\\begin{eqnarray}
\Lambda - {1\over 4} = \rho^{2}  \Lambda_{0} - {1 \over 4} \; ,
\quad
 m < 2B  , \quad  m+ n
+1/2  \leq  {eB  \over \hbar c} \rho^{2} \;  ,
\nonumber
\\
\rho^{2}  \Lambda_{0} - {1 \over 4} = 2 {eB  \over \hbar c}
\rho^{2} ( {m + \mid m \mid \over 2 } + n + 1/2)
 - (  { m + \mid
m \mid \over 2} + n + 1/2)^{2},  n =0,1,\ldots, N_{B} .
\nonumber
\label{usual}
\end{eqnarray}

\noindent  In the limit of vanishing  curvature,
 we
obtain the known result in  the flat space
$$
 E -{P^{2} \over 2M} =
 \; {eB \hbar \over M c} \; ( {m + \mid m \mid \over 2 } + n + 1/2) \; .
$$

\section{ Cox's particle in  the electric field, Minkowski space
}

Schr\"{o}dinger equation for Cox's particle in the electric field
has the form  (see Section {\bf 13})
\begin{eqnarray}
\left (  D_{t}  - c\;
 {   \Gamma ^{2} E_{i}E^{i}  \mu  +
    \Gamma    E^{j} D_{j}  \over 2( 1 +  \Gamma^{2} E_{i}E^{i}  )}  \right )     \Psi
     = -{1 \over 2M}
 \stackrel{\circ}{D}_{k}
  g^{kj}
\left [ D_{j}  + { \Gamma^{2}  E_{j} (E^{i} D_{i})   +
 \mu   \Gamma E_{j}  \over 1 + \Gamma^{2} E_{i}E^{i} }   \right  ] \Psi
\;  ;
\nonumber
\\
\label{5.1}
\end{eqnarray}

\noindent the notation is used:
\begin{eqnarray}
A_{0} = -e Ez  , \quad   E_{i}  = (F_{01} , F_{02}, F_{03})  \; ,
\quad
g^{11}E_{1} = E^{1},  \; g^{22}E_{2} = E^{2}, \; g^{33}E_{3} =
E^{3}, \; \nonumber
\\
i \hbar     \partial  _{t }    -  e  A_{0} = D_{t}  , \quad i
\hbar    \partial  _{k }    =   D_{k}  , \quad
 {i \hbar
\over \sqrt{-g}}  {\partial \over  \partial x^{k} } \sqrt{-g} =
\stackrel{\circ}{D}_{k} , \;\;\; \mu = Mc  \;. \nonumber \label{5.2}
\end{eqnarray}

\noindent Let us use cylindric coordinates
\begin{eqnarray}
dS^{2} = c^{2} dt^{2} - dr^{2} - r^{2} d\phi^{2}  - dz^{2} ,
\quad
  E_{3} =   E    ,\;
 E^{3} =   -E   ,\;  E_{3} E^{3} = - E^{2}  .
\label{5.3}
\end{eqnarray}

\noindent First, we get (let it be $\Gamma E = \gamma $)
\begin{eqnarray}
\left (  D_{t}  - c
 {   \Gamma ^{2} E_{i}E^{i}  \mu  +
     \Gamma    E^{j} D_{j}  \over 2( 1 +  \Gamma^{2} E_{i}E^{i}  )}  \right )
     =
     i\hbar \partial_{t} + e E z  +  c \; { \gamma^{2}  \mu  + \gamma
      D_{3 }   \over 2 (1 - \gamma^{2} ) }  .
 \label{5.4}
 \end{eqnarray}

\noindent Next, we consider the Hamiltonian
\begin{eqnarray}
  H =
 {1 \over 2M} \left [
 \stackrel{\circ}{D}_{1} D_{1}     +
 \stackrel{\circ}{D}_{2}    {1 \over r^{2}}
D_{2}    +
 \stackrel{\circ}{D}_{3}     \left (\;
D_{3}  + {
 \mu   \gamma   \over 1 - \gamma^{2} }   \;\right  ) \right ] .
\label{5.5}
\end{eqnarray}

\noindent
In explicit form, the  extended Schr\"{o}dinger equation  looks as
follows (to allow for the imaginary character of  $\gamma$, we
make formal change $i\gamma \longrightarrow \gamma$)
\begin{eqnarray}
\left (  i\hbar \partial_{t} + e E z  -
    { Mc^{2} \gamma^{2}  \over 2 (1 + \gamma^{2} ) } +  { \gamma \over 2(1 + \gamma^{2})}
 \hbar c \partial_{z}  \right ) \Psi
 \nonumber
 \\
=  { -\hbar^{2}  \over 2M} \left (   \partial^{2}_{r} + {1 \over
r} \partial_{r}
    +  { \partial^{2} _\phi  \over r^{2}  }
+ \partial^{2}_{z} -
 {  (Mc / \hbar)\gamma     \over 1 + \gamma^{2} }  \partial_{z}    \right ) \Psi  .
 \nonumber
  \label{5.6}
 \end{eqnarray}

With the substitution
$
\Psi = e^{-iW t / \hbar } e^{im\phi} Z(z) R(t)
$
and the notation
\begin{eqnarray}
  {M^{2} c^{2}  \over \hbar ^2} = {1 \over \lambda^{2}}  \;, \;
  {2M \over \hbar ^2}  W = w \; , \; {2M \over \hbar ^2}\; e E z = \nu  \; ,
 \nonumber
 \end{eqnarray}

 \noindent
we get
\begin{eqnarray}
     {1 \over Z(z)} \left ( \partial^{2}_{z}  + \nu \; z  + w  - {1  \over \lambda ^{2}  }
  { \gamma^{2}  \over  1 + \gamma^{2}  }        \right )  Z (z)
  + {1 \over R(r)}  \left (   \partial^{2} _{r} + {1 \over r}
\partial_{r} -  {m^{2}  \over r^{2} }\right ) R(r)  . \qquad
 \label{5.7b}
 \end{eqnarray}

\noindent After separation of the variables ($w_{\bot}>0$ stands
for the separation constant) we derive
\begin{eqnarray}
 \left (   \partial^{2} _{r} + {1 \over r}  \partial_{r} -
 {m^{2}  \over r^{2} } + w_{\bot} \right ) R(r) =  0 \;,
\nonumber
\\
 ( \partial^{2}_{z}  + \nu \; z  + w'           )  Z(z)  = 0  \; ,\qquad
 \label{5.8'}
 \end{eqnarray}
 where
 $$
  w' =  w - w_{\bot} + {1  \over \lambda ^{2} }
  { \gamma^{2}  \over  1 + \gamma^{2}  } \; .
  $$

In fact,  (\ref{5.8'}) coincide with the well
known equations
 for an ordinary particle in the uniform electric field.
Equation in the variable $z$ looks as a one-dimensional
Schr\"{o}dinger equation in  the potential of the form
 $U(z) = -
\nu\; z \;, \;\nu > 0$:
\begin{eqnarray}
( {d^{2} \over dz^{2} }    + w'     + \nu \; z       )  Z(z)  = 0
\;  . \label{5.9a'}
\end{eqnarray}

\noindent The form of the curve $U(z)$ says that any particle
moving from the right must be reflected by this barrier  in
vicinity of the point $z_{0} = - {w' \over \nu}$ (we assume that
electric force acts in positive direction of  the axis  $z$).

Mathematic solutions of the equation (\ref{5.9a'}) can be
expressed in Airy function. Indeed,  let us change
the variable
\begin{eqnarray}
\nu  z  + w'   = a x, \; ( {d^{2} \over dx^{2} }  + {a^{3} \over
\nu^{2}}  x          )  Z(x)  = 0 \; ; \nonumber
\end{eqnarray}

\noindent let it be (for definiteness  $\nu >0$)
\begin{eqnarray}
{a^{3} \over \nu^{2}}=-1 , \qquad  a = - \nu^{2/3}, \quad
 x = {\nu
 z  + w'  \over  - \nu^{2/3}}  = -\nu ^{1/3}   z - {w' \over
\nu^{2/3}}   ; \label{5.9b'}
\end{eqnarray}

\noindent then we arrive at the  Airy equation
\begin{eqnarray}
\left ( {d^{2} \over dx^{2} }   -  x     \right      )  Z(x)  = 0
\; ; \label{5.9c'}
\end{eqnarray}

\noindent to the turning point  $z_{0} = -w'/ \nu$ there
corresponds the value $x_{0}=0$.

Eq.  (\ref{5.9c'}) can be related to the Bessel equation. Indeed,
let us introduce the variable
\begin{eqnarray}
\xi = {2 \over 3} x^{3/2}\;, \qquad x ={3\over2} \xi^{2/3} \; ,
\nonumber
\label{5.10a'}
\end{eqnarray}

\noindent then Airy equation gives
\begin{eqnarray}
( {1 \over 3 \xi }  {d \over d \xi }+  {d^{2} \over d \xi^{2} }
-1) Z =0 \; . \nonumber
\end{eqnarray}

\noindent Applying the substitution $ Z = \xi^{1/3} f (\xi) $,
 we arrive at the Bessel equation  \cite{Bateman-2}
\begin{eqnarray}
\left ( {d^{2} \over d\xi^{2}} + {1 \over \xi }{d \over d \xi} -1
-{1/9 \over \xi^{2}}  \right )f (\xi)  =0  \label{5.10b'}
\end{eqnarray}

\noindent with two linearly independent solutions
\begin{eqnarray}
f_{1} (\xi) = J_{+1/3}(i \xi)\; ,  \;  f_{2} (\xi) =  J_{-1/3}(i
\xi)\;. \label{5.10c'}
\end{eqnarray}

\noindent Thus, general solutions of Airy equation can be
constructed as linear combinations of
\begin{eqnarray}
Z_{1} (x)  =  \xi^{1/3} J_{+1/3}(i \xi) \;,  \;\;  Z_{2} (x)  =
\xi^{1/3} J_{-1/3}(i \xi) \; , \nonumber
\end{eqnarray}

\noindent where
\begin{eqnarray}
 i \xi =  i {2\sqrt{\nu} \over 3}   \; (z + {w' \over \nu }) ^{3/2}  \; .
\nonumber
\label{5.11a'}
\end{eqnarray}

With the use of the known relation  \cite{Bateman-2, Bateman-2}
\begin{eqnarray}
J_{\mu} (y) = { (y/2)^{\mu} \over \Gamma (\mu +1)} \;  e^{-iy} \;
_{1}\hspace{-0.5mm} F_{1}(\mu+{1 \over2}, 2 \mu +1 , 2i y)
\nonumber
\end{eqnarray}

\noindent and with the  notation $ y = i \xi, \; \mu = +1/3, - 1
/3$,
 one expresses two independent solutions of the Schr\"{o}dinger equation as follows
\begin{eqnarray}
Z_{1}  = \xi^{1/3} J_{+1/3} (i \xi ) = \xi^{+1/3 } { (i \xi /
2)^{\mu } \over \Gamma (\mu +1)}  e^{ \xi}
 _{1}F_{1}
( +\mu  +{1 \over 2}, + 2 \mu  +1 , -2 \xi)  \;, \nonumber
\\
Z_{2}  = \xi^{1/3} J_{-1/3} (i \xi ) =  \xi^{1/3} { (i \xi /
2)^{-\mu } \over \Gamma (-\mu +1)}
 \;  e^{ \xi}
  _{1}\hspace{-0.5mm} F_{1}
(- \mu  +{1 \over 2}, -2 \mu  +1 , -2 \xi) \; . \nonumber
\end{eqnarray}

\section{ Cox's Particle in the electric field in the Lobachevsky model
}

We determine  the concept of generalized electric field  in the
special system of cylindric coordinates in the curved space as
follows \cite{Ovsiyuk-Veko}
\begin{eqnarray}
dS^{2} = dt^{2} -  \mbox{ch}^{2} z ( dr^{2} + \mbox{sh}^{2} r
d\phi^{2} ) - dz^{2}  , \nonumber
\\
\sqrt{-g} =  \mbox{sh}\; r \; \mbox{ch}^{2}z  , \; A_{0} = - E
\rho  \tanh z   , \nonumber
\\
  E_{3} =    {E \over \cosh ^{2} z } ,\; E^{3} = -
   {E  \over \cosh ^{2} z } ,
  \;
    E_{3}E^{3} = -E^{2}\;\cosh ^{-4} z   .
\label{6.1}
\end{eqnarray}

\noindent Below we use operators
 (the coordinate
$tc / \rho \longrightarrow t$ is dimensionless)
\begin{eqnarray}
i { \hbar  c \over \rho }    \partial  _{t }    -  e  A_{0} =
D_{t}  , \; i  {\hbar \over \rho}      \partial  _{k }    =
D_{k}  ,\quad
 {i \hbar / \rho  \over \sqrt{-g}}  {\partial \over
\partial x^{k} } \sqrt{-g}   = \stackrel{\circ}{D}_{k}  .
\nonumber
\end{eqnarray}

We start with the  extended Schr\"{o}dinger equation
 (let it be $E^{2}\;\cosh ^{-4} z = -\gamma^{2} (x)
$)
\begin{eqnarray}
\left (  D_{t}  - c\;
 {   \Gamma ^{2} E_{i}E^{i}  \mu  +
     \Gamma    E^{j} D_{j}  \over 2( 1 +  \Gamma^{2} E_{i}E^{i}  )}  \right )   \;  \Psi ={1 \over 2M \rho^{2}}
\nonumber
\\   =
 \stackrel{\circ}{D}_{k}     (-g^{kj})
\left ( D_{j}  + { \Gamma^{2}  E_{j} (E^{i} D_{i})   +
 \mu   \Gamma E_{j}  \over 1 + \Gamma^{2} E_{i}E^{i} }   \right  ) \Psi  .
\nonumber \label{6.2}
\end{eqnarray}

Allowing for relations
\begin{eqnarray}
\left (  D_{t}  - c\;
 {   \Gamma ^{2} E_{i}E^{i}  \mu  +
     \Gamma    E^{j} D_{j}  \over 2( 1 +  \Gamma^{2} E_{i}E^{i}  )}  \right )\qquad
\nonumber
\\
= {\hbar  c \over \rho } \left ( i  \partial_{t} + {eE\rho \over
\hbar c /  \rho }  \tanh z + {1 \over 2} {Mc \rho \over \hbar }
{ \gamma^{2}(z)   \over 1 - \gamma^{2} (z)}
 +  {1 \over 2} {\gamma  (z) \over 1 -\gamma^{2} (z) }   i  \partial_{z} \right )  ,
 \nonumber \label{6.3}
\end{eqnarray}
and
\begin{eqnarray}
H =  -{ \hbar^{2}  \over 2M  \rho^{2}}  \left [
    {1 \over \cosh^{2}z }   \left ( \partial^{2}_{r} + {\cosh r \over \sinh r} \partial_{r}   +
     {\partial^{2}_{\phi} \over \sinh ^{2} r}    \right )  +
     ( \partial_{z} + 2 { \mbox{sh}\; z \over \mbox{ch}\; z}  ) \left (  \partial_{z}
     - {Mc \rho \over \hbar }  { i \gamma (z) \over 1 - \gamma^{2} (z) }   \right )      \right ] ,
\nonumber
\end{eqnarray}

\noindent we get e an explicit form of the extended
Schr\"{o}dinger equation
\begin{eqnarray}
{\hbar  c \over \rho } \left ( i  \partial_{t} + {eE\rho \over
\hbar c /  \rho }  \tanh z + {1 \over 2} {Mc \rho \over \hbar }
{ \gamma^{2}(z)   \over 1 - \gamma^{2} (z)}
 +  {1 \over 2} {\gamma  (z) \over 1 -\gamma^{2} (z) }   i  \partial_{z} \right )
  \Psi  \qquad \qquad
\nonumber
\\
= -{ \hbar^{2}  \over 2M  \rho^{2}}
 \left [
    {1 \over \cosh^{2}z }   \left ( \partial^{2}_{r} + {\cosh r \over \sinh r} \partial_{r}   +
     {\partial^{2}_{\phi} \over \sinh ^{2} r}    \right )
          +
     ( \partial_{z} + 2 { \mbox{sh}\; z \over \mbox{ch}\; z}  ) \left (  \partial_{z}
     - {Mc \rho \over \hbar }  { i \gamma (z) \over 1 - \gamma^{2} (z) }   \right )      \right ] \Psi ;
\nonumber
\end{eqnarray}

\noindent Specially note that two terms proportional to
$i\gamma(z)  \partial_{z}$ compensate each other. Additionally, we
should perform formal change $i\gamma \longrightarrow \gamma$:
$$
{\hbar  c \over \rho } \left ( i  \partial_{t} + {eE\rho \over
\hbar c /  \rho }  \tanh z - {1 \over 2} {Mc \rho \over \hbar }
  { \gamma^{2}(z)   \over 1 + \gamma^{2} (z)}
  \right )
  \Psi
 $$
 $$
  = -{ \hbar^{2}  \over 2M  \rho^{2}}
 \left [
    {1 \over \cosh^{2}z }   \left ( \partial^{2}_{r} + {\cosh r \over \sinh r} \partial_{r}   +
     {\partial^{2}_{\phi} \over \sinh ^{2} r}    \right )  \right.
 $$
 $$
     \left. +
     ( \partial_{z} + 2 { \mbox{sh}\; z \over \mbox{ch}\; z}  )  \partial_{z}
     - {Mc \rho \over \hbar }  \left ({\partial \over \partial z}  {  \gamma (z) \over 1 + \gamma^{2} (z) } \right )
    - {Mc \rho \over \hbar }
     {  \gamma (z) \over 1+ \gamma^{2} (z) }   2 { \mbox{sh}\; z \over \mbox{ch}\; z}     \right ] \Psi ;
$$

With the use of substitution
\begin{eqnarray}
\Psi = \exp (-i w t )\;  e^{im \phi} R(r) Z(z) , \;\;  w =  {W
\rho \over \hbar c}  , \nonumber
\end{eqnarray}

\noindent and notation
\begin{eqnarray}
W =   w {\hbar c \over \rho}    {1  \over  \hbar^{2}/ 2M\rho^{2}}
= 2 w {M\rho c \over \hbar} , \quad  \nu = { e E \rho  \over
\hbar^{2}/ 2M\rho^{2}}\;, \nonumber
\\
{1 \over 2} Mc^{2} {1  \over  \hbar^{2}/ 2M\rho^{2}}= {M^{2}
\rho^{2} c^{2} \over \hbar^{2}} =\mu^{2}  ,\qquad \nonumber
\end{eqnarray}

\noindent we get
\begin{eqnarray}
 \cosh^{2}z \left (
 W  + \nu \;   \tanh z -    { \mu^{2} \gamma^{2}(z)   \over 1 + \gamma^{2} (z)}
  \right )
  RZ
+    \left ( \partial^{2}_{r} + {\cosh r \over \sinh r}
\partial_{r}   -
     {m^{2}  \over \sinh ^{2} r}    \right )  R Z
 \nonumber
 \\
      +  \cosh^{2}z \left [
     ( \partial_{z} + 2 { \mbox{sh}\; z \over \mbox{ch}\; z}  )  \partial_{z}
     - \mu  \left ({\partial \over \partial z}  {  \gamma (z) \over 1 + \gamma^{2} (z) } \right )
  - \mu   {  \gamma (z) \over 1+ \gamma^{2} (z) }   2 { \mbox{sh}\; z \over \mbox{ch}\; z}
      \right ] R Z  =0   \; .
\label{6.7}
\end{eqnarray}

In this equation one can separate the variables
\begin{eqnarray}
 \left ( {d ^{2} \over dr^{2}}  + {\cosh r \over \sinh r} {d \over dr}    -
     {m^{2}  \over \sinh ^{2} r}   + \Lambda   \right )  R =0  ,
\label{6.8}
\end{eqnarray}
\begin{eqnarray}
\left [
      {d^{2} \over dz^{2}}  + 2 { \mbox{sh}\; z \over \mbox{ch}\; z}  {d \over dz}
     - \mu  \left ({d  \over d  z}  {  \gamma (z) \over 1 + \gamma^{2} (z) } \right )
     \right. \qquad \qquad \quad
     \nonumber
     \\
     \left.
    - \mu   {  \gamma (z) \over 1+ \gamma^{2} (z) }   2 { \mbox{sh}\; z \over \mbox{ch}\; z}
       + W  + \nu \;   \tanh z
       - \mu^{2}   { \gamma^{2}(z)   \over 1 + \gamma^{2} (z)} - {\Lambda \over \mbox{ch}^{2} z}
      \right ] Z = 0 \; ;
 \label{6.9}
 \end{eqnarray}

\noindent remember that $\gamma (z) = \gamma\; \mbox{ch}^{-2} z$.
The most interesting is the equation in variable $z$. After
elementary transformation it is reduced to the form
\begin{eqnarray}
\left (
      {d^{2} \over dz^{2}}  + 2 { \mbox{sh}\; z \over \mbox{ch}\; z}  {d \over dz}\right.
      \qquad \qquad
\nonumber
\\
   \left.  - 2 \mu \gamma  \; \mbox{sh}\; z  \; \mbox{ch}\; z
     { - \mbox{ch}^{4} z + \gamma^{2} \over
     ( \mbox{ch}^{4} z  + \gamma^{2})^{2}  }
    - 2 \mu   \gamma    \;  { \mbox{sh}\; z  \; \mbox{ch}\; z   \over \mbox{ch}^{4} z  + \gamma^{2} }
        \right.
\nonumber
\\
     \left.
      + W  + \nu \;   \tanh z -   {  \mu^{2} \gamma  ^{2}  \over \mbox{ch}^{4} z  + \gamma^{2} }
         - {\Lambda \over \mbox{ch}^{2} z}
      \right ) Z) = 0 \; .
 \nonumber
 \\
 \label{6.10}
 \end{eqnarray}

This final equation turns out to be very complex and it hardly can
be solved analytically.

\section{ Solving the radial equation in the  Lobachewsky space
}

In eq.  (\ref{6.8}) let us introduce a new variable
 $x = (1+\cosh r)/2,\; x\in[1,\;+\infty)$, so that
\begin{eqnarray}
x\,(1-x)\,{d^{2}R \over dx^{2}}+(1-2\,x)\,{d R \over dx}
- \left( w_{\bot} +{1\over 4}\,{m^{2}\over x}+{1\over
4}\,{m^{2}\over 1-x}\right)\,R =0\, ; \label{7.1}
\end{eqnarray}

\noindent with the substitution $R =x^{a}\,(1-x)^{b}\,F $ at
$$
a=\pm{\mid m \mid \over 2}\,,\;\;\; b=\pm{\mid m \mid \over 2}\,
$$

\noindent  we obtain the hypergeometric equation
\begin{eqnarray}
x\,(1-x)\,{d^{2}F \over
dx^{2}}+\left[2\,a+1-(2\,a+2\,b+2)\,x\right]\,{dF \over dx}\,
-
 \left[(a+b)\,(a+b+1)+ w_{\bot} \right]\,F =0\,,
\nonumber
\end{eqnarray}

\noindent with parameters
\begin{eqnarray}
F= F (\alpha, \beta, \gamma; x) \; , \qquad \nonumber
\\
\alpha=a+b+{1\over 2}- i \sqrt{ w_{\bot} -1/4 } \; , \nonumber
\\
 \beta=a+b+{1\over 2}+ i \sqrt{ w_{\bot} -1/4}\;  ,
 \nonumber
 \\ w_{\bot} > {1 \over 4}  \, , \qquad
   \gamma=2\,a+1\,.
\label{7.2b}
\end{eqnarray}

\noindent We will specify solutions tending to zero at   $r=0$:
\begin{eqnarray}
F = u_{2} = F(\alpha ,  \beta,  \alpha + \beta +1 - \gamma,  1-x)
\; ; \label{7.3a}
\end{eqnarray}

\noindent when    $a$ and  $b$  take positive values
\begin{eqnarray}
a=+ {\mid m \mid \over 2}\,,\qquad b=+{\mid m \mid \over 2}\, ;
\label{7.3b}
\end{eqnarray}

\noindent the complete radial function is
\begin{eqnarray}
R = x ^{a} (1-x)^{b}\qquad
\times F(\alpha ,  \beta, \alpha + \beta +1 - \gamma,  1-x)  .
\label{7.3c}
\end{eqnarray}

To find behavior at infinity,  $r \rightarrow + \infty$, one
should apply the following Kummer relationship \cite{Bateman-1}
\begin{eqnarray}
u_{2} = {\Gamma(\alpha + \beta +1 - \gamma)  \Gamma(\beta -
\alpha) \over \Gamma (\beta + 1 - \gamma) \Gamma(\beta) }
e^{-i\pi \alpha}  \; u_{3} \nonumber
\\
+
        {\Gamma(\alpha + \beta +1 - \gamma ) \Gamma( \alpha - \beta)  \over
         \Gamma (\alpha +1 - \gamma) \Gamma(\alpha)} e^{-i\pi \beta} \; u_{4}\,,
\label{7.4a}
\end{eqnarray}

\noindent where
\begin{eqnarray}
u_{2} = F(\alpha ,  \beta,  \alpha + \beta +1 - \gamma;  1-x ) \;
,\qquad \nonumber
\\
u_{3} = (-x)^{-\alpha} F ( \alpha,  \alpha +1 - \gamma, \alpha +1
- \beta,  {1 \over x} )\;  , \nonumber
\\
u_{4} = (-x) ^{-\beta}  F (\beta, \beta + 1 - \gamma,   \beta +1 -
\alpha, {1 \over x} ) \; . \nonumber \label{7.4b}
\end{eqnarray}

\noindent Therefore, asymptotic  behavior at  $x \rightarrow 1 \;
(r \rightarrow + \infty)$  is given by
\begin{eqnarray}
R \approx (-1) ^{a+b}  \Gamma(\alpha + \beta +1 - \gamma)\qquad
\nonumber
\\
\times
 \left (
{  \Gamma(\beta - \alpha) \over \Gamma (\beta + 1 - \gamma)
\Gamma(\beta) }  e^{-i\pi \alpha}  \;(-x)^{a+b-\alpha}\right.
\nonumber
\\
\left.  +
        {\Gamma( \alpha - \beta)  \over
         \Gamma (\alpha +1 - \gamma) \Gamma(\alpha)} e^{-i\pi \beta} \; (-x) ^{a+b-\beta}
\right ).
     \label{7.5a}
     \end{eqnarray}

\noindent From this it follows $x \approx {e^{r} \over 4}$ and
\begin{eqnarray}
 R \approx (-1) ^{a+b}
\Gamma(\alpha + \beta +1 - \gamma)  (-x) ^{-1/2}\quad  \nonumber
\\
\times
 \left (
{  \Gamma(\beta - \alpha) \over \Gamma (\beta + 1 - \gamma)
\Gamma(\beta) }  e^{-i\pi \alpha}  \;(-x)^{+i\sqrt{\lambda -1/4} }
\right.\nonumber
\\
\left. +
  {\Gamma( \alpha - \beta)  \over
         \Gamma (\alpha +1 - \gamma) \Gamma(\alpha)} e^{-i\pi \beta} \; (-x) ^{-i\sqrt{\lambda -1/4}}
\right )\,.
  \label{7.5b}
  \end{eqnarray}
Thus, constructed solutions represent standing radial waves. The
factor $e^{-r / 2}$ is not significant for probability
interpretation
 $$
 d W = \sqrt{ -g} \; \psi^{*} \psi \; ,
$$
 and the term   $e^{-r / 2}$  will be compensated by the factor
 $\sinh r \approx e^{+r} / 2$ entering the volume element
  $
 d V = \sqrt{-g} \; dr \;dz\; d \phi.
 $

\section{Cox's particle in magnetic field in the spherical Riemann space
}

In cylindric coordinates of the spherical Riemann space (below
$\rho$ stands for the curvature radius)
\begin{eqnarray}
dS^{2} = dt^{2} - \cos^{2}z\,( dr^{2} + \sin^{2} r\, d\phi^{2})  -
dz^{2}  ,  \quad
 \sqrt{-g} = \sin\, r \,\cos^{2}z\, ,\qquad
\nonumber
\end{eqnarray}

\noindent analogue of the uniform magnetic field is given by
relations -- see  \cite{Bogush-et-al, Red'kov-Ovsiyuk}:
\begin{eqnarray}
A_{\phi} =  B \rho^{2}\, ( \cos\,r-1)  \, , \; F_{r\phi} =
  B\rho \,\sin\, r \, , \quad
\nonumber
\\
 B_{3} =   B\rho \sin  r  ,\;
 B^{3} = {B \over \rho \sin r \cos^{4}z}  ,
   B_{i}B^{i} = {B^{2}\over  \cos^{4}z} .
\nonumber \label{2.1}
\end{eqnarray}

We start with extended Schr\"{o}dinger equation in the form
\begin{eqnarray}
 \hspace{30mm}
 D_{t}    \;  \Psi =  {1 \over 2M \rho^{2}} \left [
 \stackrel{\circ}{D}_{1}   {1\over \cos^{2}z}\,    \stackrel{\ast}{D}_{1} +
 \stackrel{\circ}{D}_{2}  {1\over \sin^{2}r\,\cos^{2}z}\,    \stackrel{\ast}{D}_{2}+
 \stackrel{\circ}{D}_{3}       \stackrel{\ast}{D}_{3} \right ]\Psi \; ,
\label{2.2}
\end{eqnarray}

\noindent where
$$
D_{1} =i \hbar   \partial_{r} \;, \qquad D_{2} =i\hbar
\partial_{\phi} - {e\over c }\, B \rho^{2} ( \cos\,r-1) \;, \qquad
D_{3} =i \hbar  \partial_{z}  \;,
$$
$$
\stackrel{\circ}{D}_{1} =
 i \hbar   ( \partial_{r} + {\cos\, r \over \sin\, r} )\;,
\qquad  \stackrel{\circ}{D}_{2} = i\hbar  \partial_{\phi} -
{e\over c } B \rho^{2} ( \cos\,r-1)\;, \qquad
\stackrel{\circ}{D}_{3} =i \hbar   ( \partial_{z}  - 2 { \sin\,  z
\over \cos\,  z} ) \, ,
$$
$$
\stackrel{\ast}{D}_{1}= { (D_{1}- \Gamma B_{3}D^{2} ) \over 1 +
\Gamma^{2} B^{2}\cos^{-4} z } = {1 \over 1 + \Gamma^{2}
B^{2}\cos^{-4} z  } \left [
 i \hbar\partial_{r} + \Gamma B   \; \sin r \;\cos^{2} z
\left(i {\hbar }  \partial_{\phi} - {e\over c } B \rho^{2} (
\cos\,r-1)\right)
  \right  ]
  ,
$$
$$
\stackrel{\ast}{D}_{2} = {  ( \; D_{2} +  \Gamma B_{3} D^{1}
)\over 1  +  \Gamma^{2} B^{2} \cos^{-4} z  } =
 {1 \over 1  +  \Gamma^{2} B^{2} \cos^{-4} z } \left [
 i\hbar \partial_{\phi} - {e\over c } B \rho^{2} ( \cos\,r-1)  - i\hbar\,
  { \Gamma B \sin r \over \cos^{2}z}\,\partial_{r} \right ]
 ,
$$
$$
\stackrel{\ast}{D}_{3} = {1 \over 1 +  \Gamma^{2} B^{2}\cos^{-4} z
} (  D_{3}  +   \Gamma^{2} B^{3} \; B_{3} D_{3}  )  =
 i\hbar   \partial_{z}
\; .
$$

\noindent
With notation
$
 {eB \rho^{2} \over \hbar c } = b, \; {\Gamma B \over  \cos^{2}z} = \gamma (z)\; ,
$ we get
$$
{1 \over 2 M\rho^{2} } \stackrel{\circ}{D}_{1}  g^{11}
\stackrel{\ast}{D}_{1}
$$
$$
= {\hbar^{2} \over 2 M \rho^{2}\cos^{2}z\,
(1 + \gamma^{2}(z) ) }\,
    \left [     \partial_{r}^{2} +  {\cos r \over \sin r } \,\partial_{r}+
    i \gamma (z) b { \cos r -1 \over \sin r} \, \partial_{r}
    + {\gamma (z) \over \sin r} \partial_{r} \partial_{\phi}  - i \gamma (z)  b     \right ]  ,
$$
$$
{1 \over 2M \rho^{2} } \stackrel{\circ}{D}_{2}   g^{22}
\stackrel{\ast}{D}_{2}
$$
$$
= {\hbar^{2} \over 2M\rho^{2} \sin^{2}
r\,\cos^{2}z\, (1  + \gamma^{2}(z)) } \, \left[ \partial_{\phi} +
ib ( \cos\,r-1)\right]\, \left [
  \partial_{\phi} + ib ( \cos\,r-1)  -  \gamma(z) \sin r \, \partial_{r} \right ] ,
$$
$$
{1 \over 2M \rho ^{2} } \stackrel{\circ}{D}_{3}  g^{33}
\stackrel{\ast}{D}_{3}=
 {\hbar^{2} \over 2M\rho^{2}}\,  ( \partial_{z}  - 2 { \sin\,  z \over \cos\,  z} )\,  \partial_{z}\; .
$$

Using the substitution $ \Psi = e^{-iE t / \hbar } e^{im\phi} Z(z)
R(r) \; , \; \epsilon =
  E  / (\hbar^{2} /2 M\rho^{2} ),
$ we reduce the Schr\"{o}dinger equation to the form (the function  $\gamma (z) $ must be
imaginary:
 $i \gamma (z) \Longrightarrow \gamma (z)  $)
$$
\left [{1 \over \cos^{2}z\,(1 - \gamma^{2}(z))} \left (
\partial^{2}_{r} +  {\cos\, r  \over \sin\, r } \, \partial_{r}
  - {   [ m + b  ( \cos\,r-1)  ]^{2} \over \sin^{2} r }
   -  b \gamma (z)  \right ) \right.
   $$
   $$
   \left.  + \epsilon +  ( \partial_{z}  - 2 { \sin\,  z \over \cos\, z} )\partial_{z}
     \right ] R (r) Z(z) = 0\,.
$$

One separates the variables
\begin{eqnarray}
\hspace{30mm}\left [ {d^{2} \over dr^{2}}  +  {\cos\, r  \over
\sin\, r }  {d \over dr}
  -  {   [ m +  b  ( \cos\,r-1)  ]^{2} \over \sin^{2} r }
    +  \Lambda  \right ] R(r) = 0\,,
\label{2.8}
\end{eqnarray}
\begin{eqnarray}
\hspace{20mm} \left (
     {d^{2} \over dz^{2}}   - 2\, { \sin\,  z \over \cos\,  z} {d \over dz}
 + \epsilon
    - {b \gamma +  \Lambda  \, \cos^{2}z   \over  \cos^{4} z  - \gamma ^{2} }  \right )  Z(z) = 0 \,.
\label{2.10}
\end{eqnarray}

\section{ Analysis of the equation in the variable  $z$  }

Excluding in (\ref{2.10}) the term with first derivative, we
obtain an  equation in the  effective potential
\begin{eqnarray}
Z(z) ={1 \over  \cos\, z} f(z)\,,\; \left (
     {d^{2} \over dz^{2}}     + \epsilon +1  -U(z)
           \right )  f(z) = 0 \; ,
  \nonumber
  \\
      U(z) =  { b \gamma + \Lambda  \, \cos^{2}z   \over  \cos^{4} z  - \gamma ^{2} }\; ,
      \quad  U(z=0) = {b \gamma + \Lambda \over 1 - \gamma^{2}} \;,\quad
      \; U(z=\pm {\pi \over 2}) = -{b\over  \gamma} \; .
\label{3.1}
\end{eqnarray}

\noindent Expression for an effective force looks
\begin{eqnarray}
\hspace{20mm} F_{z} = -{d U \over dz} =  -2 \, \cos\, z  \, \sin\,
z \, {\, \Lambda \, \cos^{4} z + 2\,b\,\gamma \, \cos^{2} z
+\gamma^{2} \Lambda \over (\cos^{4} z  - \gamma ^{2})^{2} } \, ;
\label{3.2}
\end{eqnarray}

\noindent points of vanishing force (or of local extremum) are
 $
z=0\; $ and the roots of the quadratic equation
\begin{eqnarray}
\Lambda \, \cos^{4} z + 2\,b\,\gamma \, \cos^{2} z +\gamma^{2}
\Lambda =0 \qquad \Longrightarrow \qquad \left ( \cos^{2} z \right
) |_{1,2} = -{b  \over \Lambda } \gamma \pm
 \sqrt{({b^{2}  \over \Lambda ^{2} }   -  1) \gamma^{2}} \; .
\label{3.3}
\end{eqnarray}

\noindent Due to inequality   $ \Lambda^{2} > b^{2}$, under the square root is imaginary quantity.
Therefore, in physical region of the variable
 $z$ we have no other force vanishing point in addition to  $z=0$.
In the new variable $ \cos^{2}z = y$, differential equation
(\ref{2.10}) takes the form
\begin{eqnarray}
\left [ {d^{2} \over dy^{2} } +\left ( {3 \over 2} {1 \over y} +{1
\over 2}{1 \over y-1 } \right ) {d \over d  y }- { \epsilon \over
4\,y\,(y-1)} +{b\gamma+  \Lambda  \; y   \over  (y   - \gamma)\,
(y+ \gamma )\,  4 \,y\,(y-1)} \right ] Z(y) = 0 \; . \label{3.5}
\end{eqnarray}

\noindent Behavior near the five singular pints can be found
straightforwardly.

\vspace{3mm}

$y \sim 1 \;\; (z \rightarrow 0)$

\begin{eqnarray}
\hspace{20mm} \left [ {d^{2} \over dy^{2} } +\left ( {1 \over 2}{1
\over y-1 } \right ) {d \over y }- { \epsilon \over 4(y-1)} +
{b\gamma+ \Lambda    \over  (1   - \gamma^{2})  \, 4 \,(y-1)}
\right ] Z(y) = 0\; , \nonumber
\\
\hspace{20mm} Z(y)=\exp \,[ \pm  \sqrt{A (y-1)}]   , \qquad  A =
\epsilon - {b \gamma +\Lambda \over 1 - \gamma^{2}} \; ;
\label{3.6}
\end{eqnarray}

$ y \sim 0$
\begin{eqnarray}
\hspace{20mm} \left ( {d^{2} \over dy^{2} } + {3 \over 2 y} \,{d
\over dy } + { \epsilon \over 4y }  +  {b \over    4 \gamma    y }
\right ) Z(y) = 0 \; ,
\nonumber
\\
 Z(y)= { \exp [ \; \pm  \sqrt{ C y}\;
]\over \sqrt{y}} , \quad C = -\epsilon - {b \over \gamma} \; ;
\label{3.7}
\end{eqnarray}

$ y \sim \infty$
\begin{eqnarray}
\hspace{20mm} \left (  {d^{2} \over  dy^{2}} + {2 \over y} {d
\over dy} -
 {\epsilon  \over 4 y^{2}} \right  )   Z(z) = 0 , \quad Z = y^{D} \; , \quad
D={-1 \pm  \sqrt{\epsilon +1 } \over 2} \,; \label{3.8}
\end{eqnarray}

$ y \sim + \gamma $
\begin{eqnarray}
\hspace{20mm} \left [ {d^{2} \over dy^{2} } + { 1\over 2} \left (
{3 \over \gamma} + {1 \over \gamma -1 } \right ) {d \over d y } +
{ \Lambda  +b  \over   8 \gamma ( 1 -\gamma )} \;\; {1 \over y   -
\gamma}\right ] Z(y) = 0 \nonumber
\end{eqnarray}

\noindent or
\begin{eqnarray}
\left [ (y - \gamma) {d^{2} \over dy^{2} } + M  (y-\gamma) {d
\over d y } +  N \right ] Z(y) = 0 ,\quad M={ 1\over 2} \left (
{3 \over \gamma} + {1 \over \gamma -1 } \right )\,,\quad N={
\Lambda  +b  \over   8 \gamma ( 1 -\gamma )}\,. \nonumber
\end{eqnarray}
Changing the variable $ -M  (y-\gamma ) = x$ we get
 \begin{eqnarray}
 \hspace{30mm} \left ( x {d^{2} \over dx^{2}} - x {d \over dx} - \alpha \right ) Z =0,
\qquad \alpha = {N \over M} \; , \nonumber
\end{eqnarray}

\noindent which is confluent hypergeometric equation of a special
form
\begin{eqnarray}
\hspace{30mm} \left ( x {d^{2} \over dx^{2}} + (c-x)  {d \over dx}
-a \right ) Z =0 \; , \qquad c=0, \;\;\; a = {N \over M}= {
\Lambda + b  \over 4(3- 4\gamma )} \; ; \nonumber \label{3.9b}
\end{eqnarray}

\noindent its general solution looks as $ Z = c_{1} M(a+1, 2, y) +
c_{2} U(a+1,2, y).$

Now, let us consider the case $ y \sim - \gamma $:
\begin{eqnarray}
\hspace{30mm} \left [ {d^{2} \over dy^{2} }  {1 \over 2} \left (
{3 \over -\gamma} +
 {1 \over  -\gamma -1 } \right ) {d \over d y } +
 { \Lambda  -b     \over   8   (- \gamma ) ( 1 +  \gamma )} \; {1 \over y+ \gamma }\right ] Z(y) = 0
\nonumber
\end{eqnarray}
or shorter
\begin{eqnarray}
\left [ (y - \gamma) {d^{2} \over dy^{2} } + M ' (y-\gamma) {d
\over d y } +  N '\right ] Z(y) = 0 , \; -M'  (y+\gamma ) = x \;
,
\nonumber
\\
 \left ( x {d^{2} \over dx^{2}} - x {d \over dx} - \alpha
' \right ) Z =0, \; a' = {N' \over M'} \; , \nonumber
\end{eqnarray}

\noindent which is confluent hypergeometric equation of a special
form
\begin{eqnarray}
\hspace{30mm} \left ( x {d^{2} \over dx^{2}} + (c-x)  {d \over dx}
-a \right ) Z =0 =0 \; , \; c=0, \;\;\; \alpha
 = {N' \over M'}=
{ \Lambda  - b \over 4(3+ 4\gamma )}  \;; \nonumber
\end{eqnarray}

\noindent its general solution is $ Z = c_{1} M(a+1, 2, y) + c_{2}
U(a+1,2, y) $.

Further analytical treatment of the differential equation
(\ref{3.5}) is very difficult because  complexity of this
equation.

\section{Solutions of the radial equation
}

Now let us consider the radial equation
 (\ref{2.8}).
 It can be solved in hypergeometric  functions -- see \cite{Bogush-et-al, Red'kov-Ovsiyuk}.
 Below we will write down only energy spectrum.
In usual measure units these formulas read
\begin{eqnarray}
  m > 0 ,
\quad \rho^{2}\Lambda_{0} +{1 \over 4} = + 2{eB  \over \hbar c}\rho^{2}
(n+m+1/2) +(n+m+1/2)^{2} \; ;
\nonumber
\\
  m < -2 {eB  \over \hbar c} \; \rho^{2}, \quad
  \rho^{2}\Lambda_{0}  +{1 \over 4} = -2 {eB  \over \hbar c}\rho^{2} (n-m+1/2) +(n-m+1/2)^{2} \; ;
\nonumber
\\
 -2{eB  \over \hbar c} \; \rho^{2} < m
\leq 0\; ,  \quad
  \rho^{2}\Lambda_{0} +{1 \over 4} = 2 {eB  \over \hbar c} \rho^{2}(n+1/2) +(n+1/2)^{2} \; .
\end{eqnarray}

Transition to the limit of the Minkowski space is achieved
accordingly to ( $\rho \rightarrow \infty$)
\begin{eqnarray}
 m < 0\; ,   \qquad
  \Lambda_{0}  = 2 {eB  \over \hbar c} (n+1/2)  \,; \qquad
m \geq 0 \; ,  \qquad  \Lambda_{0} = + 2{eB
\over \hbar c} (n+m+1/2)  \,, \label{4.15}
\end{eqnarray}

\noindent Thus, we get the well-known result
\begin{eqnarray}
 E -{P^{2} \over 2M} =
 \; {eB \hbar \over M c} \; ( {m + \mid m \mid \over 2 } + n + 1/2) \; .
\label{4.16}
\end{eqnarray}

\section{Cox's particle in electric field in the spherical model
}

In cylindrical coordinate
$
dS^{2} = dt^{2} - \cos^{2}z\,( dr^{2} + \sin^{2} r\, d\phi^{2})  -
dz^{2}  ,
$
external electric  field along the axis $z$   is given by
\begin{eqnarray}
A_{0} = - E \,\rho\,  \tan z  \, , \;\;\; E_{3} =
   {E \over \cos ^{2} z } \,,
   \quad
    E^{3} = -
   {E  \over \cos ^{2} z }\, ,
\;\;
    E_{3}\,E^{3} = -{E^{2}\over\cos ^{4} z } \, .
\label{6.1}
\end{eqnarray}

\noindent Below we use operators (and dimensionless coordinate $tc
/ \rho \longrightarrow t$)
\begin{eqnarray}
i { \hbar  c \over \rho }    \partial  _{t }    -  e  A_{0} =
D_{t}  , \; i  {\hbar \over \rho}      \partial  _{k } =   D_{k}
,
 {i \hbar / \rho  \over \sqrt{-g}}
{\partial \over  \partial x^{k} } \sqrt{-g}   =
\stackrel{\circ}{D}_{k}  . \nonumber
\end{eqnarray}

We start with the  form of Schr\"{o}dinger equation
\begin{eqnarray}
\left (  D_{t}  - c\;
 {   \Gamma ^{2} E_{i}E^{i}  \mu  +
     \Gamma    E^{j} D_{j}  \over 2( 1 +  \Gamma^{2} E_{i}E^{i}  )}  \right )     \Psi
     = {
 \stackrel{\circ}{D}_{k}     (-g^{kj})  \over 2M \rho^{2}}
\left ( D_{j}  + { \Gamma^{2}  E_{j} (E^{i} D_{i})   +
 \mu   \Gamma E_{j}  \over 1 + \Gamma^{2} E_{i}E^{i} }   \right  ) \Psi  .
\label{6.2}
\end{eqnarray}

\noindent After needed calculation
 we get representation for the wave equation
$$
{\hbar  c \over \rho } \left ( i  \partial_{t} + {eE\rho \over
\hbar c /  \rho }  \tan z + {1 \over 2} {Mc \rho \over \hbar }   {
\gamma^{2}(z)   \over 1 - \gamma^{2} (z)}
 +  {1 \over 2} {\gamma  (z) \over 1 -\gamma^{2} (z) }   i  \partial_{z} \right )
  \Psi
  $$
 $$
   = -{ \hbar^{2}  \over 2M  \rho^{2}}
 \left [
    {1 \over \cos^{2}z }   \left ( \partial^{2}_{r} + {\cos r \over \sin r} \partial_{r}   +
     {\partial^{2}_{\phi} \over \sin ^{2} r}    \right )  +
   \left( \partial_{z} - 2 \,{ \sin  z \over \cos  z} \right ) \left ( {1-2 \gamma^{2}(z)\over 1 - \gamma^{2}(z)}
    \partial_{z}
     - {Mc \rho \over \hbar }  { i \gamma (z) \over 1 - \gamma^{2} (z) }   \right )    \right ] \Psi .
$$

\noindent Note that two terms proportional to $i\gamma(z)
\partial_{z}$ compensate each other; besides we make  the formal change
 $i\gamma \longrightarrow \gamma$:
\begin{eqnarray}
{\hbar  c \over \rho } \left ( i  \partial_{t} + {eE\rho \over
\hbar c /  \rho }  \tan z - {1 \over 2} {Mc \rho \over \hbar }
  { \gamma^{2}(z)   \over 1 + \gamma^{2} (z)}
  \right )
  \Psi \hspace{20mm}
  \nonumber
  \\
  = -{ \hbar^{2}  \over 2M  \rho^{2}}
 \left [
    {1 \over \cos^{2}z }   \left ( \partial^{2}_{r} + {\cos r \over \sin r} \,\partial_{r}   +
     {\partial^{2}_{\phi} \over \sin ^{2} r}    \right )  +
    \left ( \partial_{z} - 2\, { \sin\, z \over \cos\, z} \right )
     \, \left ( {1+2 \,\gamma^{2}(z)\over 1 + \gamma^{2}(z)}\, \partial_{z}\right)\right.
    \nonumber
    \\
     \left.
     - {Mc \rho \over \hbar }  \,{\partial \over \partial z}  {  \gamma (z) \over 1 + \gamma^{2} (z) }
    + {Mc \rho \over \hbar }  {  \gamma (z) \over 1+ \gamma^{2} (z) }
     \, 2\, { \sin\,z \over \cos\, z}     \right ] \Psi . \hspace{20mm}
\label{6.5}
\end{eqnarray}

With substitution $ \Psi =e^{-i w t }\;  e^{im \phi} \,R(r)\, Z(z)
\,, \;  w =  {W \rho \over \hbar c}$, and notation
\begin{eqnarray}
W =   w {\hbar c \over \rho}  \;  {1  \over  \hbar^{2}/
2M\rho^{2}} = 2 w {M\rho c \over \hbar} \;,\quad  \nu = e E
\rho\; {1  \over  \hbar^{2}/ 2M\rho^{2}}\;,\quad {1 \over 2}
Mc^{2}\; {1  \over  \hbar^{2}/ 2M\rho^{2}}= {M^{2} \rho^{2} c^{2}
\over \hbar^{2}}\; =\mu^{2} \; , \nonumber
\end{eqnarray}

\noindent we arrive at
\begin{eqnarray}
 \cos^{2}z \left (
 W  + \nu \;   \tan z - \mu^{2}   { \gamma^{2}(z)   \over 1 + \gamma^{2} (z)}
  \right )
  R(r) Z(z) \qquad \qquad \qquad
  \nonumber
  \\
  +
     \left ( \partial^{2}_{r} + {\cos r \over \sin r} \,\partial_{r}   -
     {m^{2}  \over \sin ^{2} r}    \right )  R(r) Z(z)  +  \cos^{2}z \left [
      \left ( \partial_{z} - 2\, { \sin\, z \over \cos\, z} \right )  \,
      \left ( {1+2 \,\gamma^{2}(z)\over 1 + \gamma^{2}(z)}\, \partial_{z}\right)\right.
 \nonumber
 \\
   \left.  - \mu \,{\partial \over \partial z} \, {  \gamma (z) \over 1 + \gamma^{2} (z) }
    + \mu   {  \gamma (z) \over 1+ \gamma^{2} (z) }  \, 2\, { \sin\, z \over \cos\, z}
      \right ] R(r) Z(z)  =0 \,  . \qquad \qquad \qquad
\label{6.7}
\end{eqnarray}

After the separation of the variable we obtain
$$
 \left ( {d ^{2} \over dr^{2}}  + {\cos r \over \sin r} {d \over dr}    -
     {m^{2}  \over \sin ^{2} r}   + \Lambda   \right )  R(r) =0 \; ,
\eqno(6.8)
$$

\noindent this equation can be readily solved in hypergeometric functions. In turn, equation in $z$ variable is
\begin{eqnarray}
 \left [
      \left ( {d \over d z} - 2\, { \sin\, z \over \cos\, z} \right )  \, \left ( {1+2 \,\gamma^{2}(z)\over 1 + \gamma^{2}(z)}\, {d \over d z}\right)
      - \mu \,{d \over d z} \, {  \gamma (z) \over 1 + \gamma^{2} (z) }
    + \mu   {  \gamma (z) \over 1+ \gamma^{2} (z) }  \, 2\, { \sin\, z \over \cos\, z}
        \right.
\nonumber
\\
     \left.
      + W  + \nu \,   \tan z - \mu^{2}   { \gamma^{2}(z)   \over 1 + \gamma^{2} (z)} - {\Lambda \over \cos^{2} z}
      \right ] Z(z) = 0 \; . \qquad\qquad \qquad
 \label{6.9}
 \end{eqnarray}

\noindent Remember that $ \gamma (z) = \gamma\,\cos^{-2} z . $ The
last equation can be translated to the following form
\begin{eqnarray}
\left (
      {\cos^{4}z+2\,\gamma^{2}\over \cos^{4}z+\gamma^{2}}\,{d^{2} \over dz^{2}}  - 2 \,{ \sin\, z \over \cos\, z}  \,{\gamma^{2}\,\cos^{4}z+2\,\gamma^{4}+\cos^{8}z\over (\cos^{4}z+\gamma^{2})^{2}}\,{d \over dz}-\mu\,\gamma\,{\cos^{2}z\over \cos^{4}z+\gamma^{2}}\,{d\over dz}
     \right.
 \nonumber
 \\
     \left.
      + 4 \,\mu\, \gamma^{3}  \, {\sin\, z  \, \cos\, z
     \over
     ( \cos^{4} z  + \gamma^{2})^{2}  }
   +W  + \nu \,   \tan z -   {  \mu^{2} \gamma  ^{2}  \over \cos^{4} z  + \gamma^{2} }
         - {\Lambda \over \cos^{2} z}
      \right ) Z(z) = 0 \; .
 \label{6.10}
 \end{eqnarray}

We could not proceed further with this differential equation
because of its complexity.

\section{Cox's particle in arbitrary curved space time, general analysis}

First, we  will use a Proca-like generalized system  in  Cartesian coordinates in  Minkowski space
   (note  the
notation:   $\mu = Mc$)
\begin{eqnarray}
 (\mu\delta_{\alpha}^{\beta}+\lambda
F_{\alpha}^{\;\;\beta})\Phi_{\beta}=D_{\alpha}\Phi\; , \;
D^{\alpha}\Phi_{\alpha}=\mu\Phi \; ,
\nonumber
\label{1.1} \end{eqnarray}

\noindent or shorter
\begin{eqnarray}
\Lambda_{\alpha}^{\;\;\beta}\Phi_{\beta}=D_{\alpha}\Phi\; , \;
D^{\alpha}\Phi_{\alpha}=\mu\Phi \; . \label{1.2}
\end{eqnarray}

The first equation in (\ref{1.2}) can be multiplied by the inverse
matrix $(\Lambda^{-1})_{\rho}^{\;\;\alpha}$, so we get
\begin{eqnarray}
 \Phi_{\rho} =
(\Lambda^{-1})_{\rho}^{\;\;\alpha}D_{\alpha}\Phi \; , \; D^{\rho}
\Phi_{\rho}=\mu \Phi \; . \label{1.3}
\end{eqnarray}

\noindent
 From (\ref{1.3}), one  can derive a generalized
Klein--Fock--Gordon equation for the scalar function $\Phi$:
\begin{eqnarray}
\left [\;  \mu D^{\rho}(\Lambda^{-1})_{\rho}^{\;\;
\alpha}D_{\alpha} \; - \; \mu^{2} \right ]  \Phi=0\; .
\label{1.4a}
\end{eqnarray}

\noindent Equation (\ref{1.4a}) can be rewritten  in the form
\begin{eqnarray}
\{  \mu (\Lambda^{-1})_{\rho}^{\;\; \alpha} \; D^{\rho} D_{\alpha}
+ \mu\;  [\; i\hbar \partial ^{\rho} (\Lambda^{-1})_{\rho}^{\;\;
\alpha} ] \; D_{\alpha}
 - \mu^{2}  \}  \Phi =0 \; .
\label{1.4b}
\end{eqnarray}

We are to find an  explicit form
of the inverse matrix
 $\Lambda^{-1}$ (the calculation
of the inverse matrix performed in this section is valid only in Cartesian coordinates of the
flat space;  generalization to the case of a curved space
 or curvilinear coordinates in the flat space will be given below):
\begin{eqnarray}
 \Lambda= (\Lambda_{\alpha}^{\;\; \beta}) = \left |
\begin{array}{rrrr}
\mu & - e_{1} & - e_{2} &- e_{3} \\
- e_{1} & \mu & - b_{3} &  b_{2} \\
- e_{2}&  b_{3} & \mu & - b_{1} \\
- e_{3}& -  b_{2} &  b_{1} & \mu
 \end{array} \right |,\quad
  e_{i} = \lambda E_{i}, \; b_{i}= \lambda B_{i}.
\label{1.5}
\end{eqnarray}

 The
inverse matrix is defined by the  formula
\begin{eqnarray}
\Lambda^{-1}= {1 \over \det \Lambda } \left | \begin{array}{rrrr}
M_{0}^{\;\;0}  &  - M_{1}^{\;\;0}  & +M_{2}^{\;\;0}  & - M_{3}^{\;\;0}\\
-M_{0}^{\;\;1}  &  + M_{1}^{\;\;1}  & -M_{2}^{\;\;1}  & + M_{3}^{\;\;1}\\
+M_{0}^{\;\;2}  &  - M_{1}^{\;\;2}  & +M_{2}^{\;\;2}  & - M_{3}^{\;\;2}\\
-M_{0}^{\;\;3}  &  + M_{1}^{\;\;3}  & -M_{2}^{\;\;3}  & +
M_{3}^{\;\;3}
\end{array} \right | .
\nonumber
\end{eqnarray}

\noindent The determinant of the  matrix $\Lambda$ equals to
$ \det \Lambda = \mu^{4}  - \mu^{2} \;   ( \vec{e}^{\;2} -
\vec{b} ^{\;2})   -
 ( \vec{e} \;  \vec{b})^{2}$ ,
   and   the  minors are
$$
M_{0}^{\;\;0} = \mu^{3}  +  \mu  \; \vec{b}^{\;2}\; , \quad
M_{1}^{\;\;1} =  \mu^{3}+ \mu (b_{1}^{2}-e_{2}^{2}-e_{3}^{2}) \; ,
$$
$$
M_{2}^{\;\;2} =  \mu^{3} + \mu (b_{2}^{2}-e_{1}^{2}-e_{3}^{2}) \;
, \quad
 M_{3}^{\;\;3} = \mu^{3} + \mu (b_{3}^{2}-e_{1}^{2}-e_{2}^{2})\;
,
$$
$$
M_{0}^{\;\;1} =  - \mu^{2}   e_{1}
  - \mu  ( e_{2} b_{3}- e_{3}b_{2} ) -   b_{1} (\vec{e}\; \vec{b}) \; ,
\quad
M_{1}^{\;\;0} = - \mu^{2}  e_{1}
  + \mu (e_{2} b_{3}-e_{3}b_{2} ) -  b_{1} (\vec{e}\; \vec{b})\; ,
$$
$$
M_{0}^{\;\;2} =\mu^{2} e_{2}
  + \mu (e_{3}b_{1} - e_{1} b_{3}) +  b_{2} (\vec{e}\; \vec{b})\; ,
\quad
 M_{2}^{\;\;0} = \mu^{2} e_{2}
  - \mu (e_{3}b_{1} - e_{1} b_{3}) +  b_{2} (\vec{e}\; \vec{b})\; ,
$$
$$
M_{0}^{\;\;3}=-\mu^{2}  e_{3}
  - \mu (  e_{1} b_{2} -e_{2}b_{1} ) -  b_{3} (\vec{e}\; \vec{b})\; ,
\quad
 M_{3}^{\;\;0} =-\mu^{2} e_{3}
  + \mu (e_{1} b_{2}-e_{2} b_{1} )-  b_{3} (\vec{e}\; \vec{b}) \; ,
$$
$$
M_{1}^{\;\;2} =  \mu^{2} b_{3}
  - \mu (e_{1} e_{2}+b_{1}b_{2} ) -   e_{3} (\vec{e}\; \vec{b})\; ,
\quad
 M_{2}^{\;\;1}= -\mu^{2} b_{3}
  - \mu (e_{1} e_{2}+b_{1}b_{2} )+   e_{3} (\vec{e}\; \vec{b})\; ,
$$
$$
M_{1}^{\;\;3} =\mu^{2}b_{2}
  + \mu (e_{1} e_{3}+b_{1}b_{3} ) -  e_{2} (\vec{e}\; \vec{b})\; ,
\quad
M_{3}^{\;\;1} =-\mu^{2} b_{2}
  + \mu (e_{1} e_{3}+b_{1}b_{3} )+   e_{2} (\vec{e}\; \vec{b})\; ,
$$
$$
M_{2}^{\;\;3} = \mu^{2}  b_{1}
  - \mu (e_{2} e_{3}+b_{2}b_{3} )-  e_{1} (\vec{e}\; \vec{b})\; ,
\quad
 M_{3}^{\;\;2} = -\mu^{2} b_{1}
  - \mu (e_{2} e_{3}+b_{2}b_{3} )+  e_{1} (\vec{e}\; \vec{b})\; .
$$

\noindent Let us detail two  simple  cases. The
first is the presence of the electric field:
\begin{eqnarray}
B_{i}=0\;, \qquad \Lambda=   \left |
\begin{array}{rrrr}
\mu & - e_{1} & - e_{2} &- e_{3} \\
- e_{1} & \mu & 0 &  0 \\
- e_{2}&  0 & \mu & 0 \\
- e_{3}& -  0 &  0 & \mu
 \end{array} \right |,\nonumber
 \end{eqnarray}
 \begin{eqnarray}
\; \Lambda^{-1} = {1 \over \mu^{4}- \mu^{2} \vec{e}^{\;2}}
\left |
\begin{array}{rrrr}
\mu^{3}  &  \mu^{2}  e_{1}  & \mu^{2} e_{2}  & \mu^{2} e_{3} \\
\mu^{2} e_{1}  & \mu^{3}- \mu (e_{2}^{2} +e_{3}^{2})  & \mu e_{1}e_{2} &   \mu e_{1}e_{3}\\
\mu^{2} e_{2} &  \mu  e_{1}e_{2} & \mu^{3}  - \mu  ( e_{1}^{2} +e_{3}^{2})  &   \mu  e_{2}e_{3}\\
 \mu^{2}  e_{3} &  \mu e_{1}e_{3}  &  \mu  e_{1}e_{3} & \mu^{3}  - \mu( e_{1}^{2} +e_{3}^{2})
\end{array} \right |  .
\nonumber
\label{1.8}
\end{eqnarray}

\noindent
The second is the case of the magnetic field:
\begin{eqnarray}
\hspace{-8mm} E_{i}=0 ,  \Lambda= \left | \begin{array}{rrrr}
\mu & 0 & 0 &0 \\
0 & \mu & -  b_{3} &  b_{2} \\
0&   b_{3} & \mu & - b_{1} \\
0& -  b_{2} &   b_{1} & \mu
 \end{array} \right |,\qquad \Lambda^{-1} = {1 \over \mu^{4} + \mu^{2}  \vec{b}^{\;2} }
 \nonumber
 \\
 \times
 \left | \begin{array}{cccc}
\mu^{3} +\mu \vec{b}^{\;2}  & 0  & 0  & 0 \\
0 & \mu^{3} + \mu b_{1}^{2}   & \mu^{2} b_{3}+\mu b_{1}b_{2} &  -\mu^{2} b_{2}+\mu b_{1}b_{3}\\
0 & -\mu^{2} b_{3}+\mu b_{1}b_{2} & \mu^{3} + \mu b_{2}^{2}   &   \mu^{2} b_{1}+\mu b_{2}b_{3}\\
0 &    \mu^{2} b_{2}+\mu b_{1}b_{3} & -\mu^{2} b_{1}+\mu
b_{2}b_{3} & \mu^{3} + \mu b_{3}^{2}
\end{array} \right | .
\nonumber
\label{1.9}
\end{eqnarray}

Now, let us derive an explicit form of the generalized scalar equation
in the case of the  uniform magnetic field:
\begin{eqnarray}
 K ^{\rho \alpha}=  g^{\rho \beta} ( \Lambda^{-1} )_{\beta}^{\;\;\alpha} =
{1 \over \mu^{4} + \mu^{2} \vec{b}^{\;2}} \qquad\qquad\qquad\qquad
\nonumber
\\
\times \left |
\begin{array}{rrrr}
\mu^{3} +\mu \vec{b}^{\;2}  & 0  & 0  & 0 \\
0 & -\mu^{3} - \mu b_{1}^{2}   & -\mu^{2} b_{3}-\mu b_{1}b_{2} &  \mu^{2} b_{2}-\mu b_{1}b_{3}\\
0 & \mu^{2} b_{3}- \mu b_{1}b_{2} & -\mu^{3}  - \mu b_{2}^{2}   &   - \mu^{2} b_{1}- \mu b_{2}b_{3}\\
0 &   -  \mu^{2} b_{2}- \mu  b_{1}b_{3} & \mu^{2} b_{1} -\mu
b_{2}b_{3} & - m^{3}- \mu b_{3}^{2}
\end{array} \right | .
\label{2.1}
\end{eqnarray}

\noindent We are to specify the  equation (\ref{1.4a})
\begin{eqnarray}
\left (  K ^{\rho \alpha} D_{\rho} D_{\alpha}     - \mu
\right ) \Phi =0 \; , \qquad
 ( D_{\alpha} ) = ( i\hbar
\partial_{0} , \; i\hbar \partial_{j} -  {e \over  c}  A_{j})
\label{2.2}
\end{eqnarray}
in the field $ \vec{A} = {1 \over 2}  \vec{x} \times \vec{B} $.
After simple calculation we arrive at (for brevity we use  the
 parameter $\lambda / \mu= \Gamma$)
\begin{eqnarray}
\left  [\;   ( 1 +  \Gamma^{2} \vec{B}^{\;2} )   D^{2}_{0}   -
\vec{D}^{\;2} + i\hbar {e \over c}
 \Gamma  \vec{B}^{\;2}   -
  \Gamma^{2}  ( \vec{B} \vec{D} ) ^{2}-
 \mu^{2}
 ( 1 + \Gamma^{2}  \vec{B}^{\;2}) \; \right   ]  \Phi = 0 \; .
\label{2.4}
\end{eqnarray}

Now  let us consider the case of homogeneous electric field:
\begin{eqnarray}
 A_{0} =  -\vec{E}\; \vec{x}
, \;
 \bar{K} ^{\rho \alpha}=
 {1 \over \mu^{4}  - \mu^{2}  \vec{e}^{\;2}} \qquad\qquad\quad
 \nonumber
 \\
 \times
 \left | \begin{array}{rrrr}
\mu^{3}   &  \mu^{2} e_{1}  & \mu^{2}  e_{2}  & \mu^{2}  e_{3} \\
-\mu^{2} e_{1}  & -\mu^{3}  + \mu (e_{2}^{2} +e_{3}^{2})  & -\mu e_{1}e_{2} &   -\mu e_{1}e_{3}\\
-\mu^{2} e_{2} &  -\mu e_{1}e_{2} & -\mu^{3}  + \mu ( e_{1}^{2} +e_{3}^{2})  &   -\mu e_{2}e_{3}\\
- \mu^{2} e_{3} &  -\mu  e_{1}e_{3}  &  -\mu  e_{2}e_{3} &
-\mu^{3}  + \mu ( e_{1}^{2} +e_{2}^{2})
\end{array} \right |  .
\label{3.1}
\end{eqnarray}

\noindent The  equation (\ref{1.4a})
will take the form
\begin{eqnarray}
\left [ \;    D^{2}_{0}
   -( 1 -\Gamma^{2}  \vec{E}^{\;2})  \vec{D}^{\;2}
      - i \hbar{ e  \over c} \Gamma  \vec{E}^{\;2}
      -
    \Gamma^{2}  ( \vec{E} \vec{D} )^{2}
   -    \mu^{2}  (  1 - \Gamma^{2} \vec{E}^{\;2})\;
 \right ] \Phi =0 \; . \label{3.3}
\end{eqnarray}

Let us construct the inverse matrix $\Lambda^{-1}$ in   tensor notation.
 We start from the explicit form of the tensor
$(F_{\alpha} ^{\;\; \beta})$:
\begin{eqnarray}
(F_{\alpha} ^{\;\; \beta}) = \left | \begin{array}{rrrr}
0 & -E_{1} & -E_{2} &-E_{3} \\
-E_{1} & 0 & -B_{3} & B_{2} \\
-E_{2}& B_{3} & 0 & -B_{1} \\
-E_{3}& -B_{2} &  B_{1} & 0
 \end{array} \right | ,
 \label{4.1}
 \end{eqnarray}

\noindent then compute the convolution of two tensors (in this
section we perform calculations, based on Cartesian coordinates in
Minkowski space)
\begin{eqnarray}
(F_{\alpha} ^{\;\; \beta}) (F_{\beta} ^{\;\; \rho})= \left |
\begin{array}{rrrr}
\vec{E}^{\;2} & -E_{2}B_{3}+E_{3}B_{2} & E_{1}B_{3}-E_{3}B_{1} & -E_{1}B_{2}+E_{2}B_{1} \\
E_{2}B_{3}-E_{3}B_{2} & E_{1}^{2}-B_{2}^{2}-B_{3}^{2} & E_{1}E_{2}+B_{1}B_{2} & E_{1}E_{3}+B_{1}B_{3} \\
-E_{1}B_{3}+E_{3}B_{1}& E_{1}E_{2}+B_{1}B_{2} &  E_{2}^{2}-B_{1}^{2}-B_{3}^{2} & E_{2}E_{3}+B_{2}B_{3} \\
E_{1}B_{2}-E_{2}B_{1}& E_{1}E_{3}+B_{1}B_{3} &
E_{2}E_{3}+B_{2}B_{3} &  E_{3}^{2}-B_{1}^{2}-B_{2}^{2}
 \end{array} \right |.
\label{4.2}
\end{eqnarray}

\noindent Compute the convolution of three tensors:
\begin{eqnarray}
(F_{\alpha} ^{\;\; \beta}) (F_{\beta} ^{\;\; \rho})(F_{\rho}
^{\;\; \sigma})= (\vec{E}^{\;2}-\vec{B}^{\;2}) \left |
\begin{array}{rrrr}
0 & -E_{1} & -E_{2} &-E_{3} \\
-E_{1} & 0 & -B_{3} & B_{2} \\
-E_{2}& B_{3} & 0 & -B_{1} \\
-E_{3}& -B_{2} &  B_{1} & 0
 \end{array} \right |
 \nonumber
 \\
 +
(\vec{E}\vec{B}) \left | \begin{array}{rrrr}
0 & -B_{1} & -B_{2} &-B_{3} \\
-B_{1} & 0 & E_{3} & -E_{2} \\
-B_{2}& -E_{3} & 0 & E_{1} \\
-B_{3}& E_{2} &  -E_{1} & 0
 \end{array} \right |. \qquad\qquad
\label{4.3}
\end{eqnarray}

\noindent
Next compute the convolution of the electromagnetic tensor and
tensor dual to it:
\begin{eqnarray}
 (F_{\alpha} ^{\times  \beta}) (F_{\beta} ^{\;\; \rho})
= \left | \begin{array}{rrrr}
\vec{E}\vec{B} & 0 & 0 & 0 \\
0 & \vec{E}\vec{B} & 0 & 0 \\
0 & 0 & \vec{E}\vec{B}& 0 \\
0 & 0 &  0 & \vec{E}\vec{B}
 \end{array} \right | .
\label{4.4}
\end{eqnarray}

\noindent
Using (\ref{4.3}) and (\ref{4.4}), we find the explicit form of
the convolution of four tensors:
\begin{eqnarray}
\hspace{30mm}
(F_{\alpha} ^{\;\; \beta}) (F_{\beta} ^{\;\; \chi})(F_{\chi}
^{\;\; \sigma})(F_{\sigma} ^{\;\; \rho})=
(\vec{E}^{\;2}-\vec{B}^{\;2})(F_{\alpha} ^{\;\; \beta}) (F_{\beta}
^{\;\; \rho})+(\vec{E}\vec{B})^{\;2}\delta_{\alpha}^{\;\;\rho} \;
. \label{4.5}
\end{eqnarray}

\noindent Relationship  (\ref{4.5}) provides us with the
minimal polynomial of the electromagnetic matrix $(F_{\alpha}
^{\;\; \beta})$. The inverse of the $
\Lambda_{\sigma}^{\;\;\alpha}= \mu
\delta_{\sigma}^{\;\;\alpha}+\lambda F_{\sigma}^{\;\;\alpha} $,
the matrix $(\Lambda^{-1})_{\alpha}^{\;\;\beta}$,  is to be searched
in the form the following linear combination
\begin{eqnarray}
 (\Lambda^{-1})_{\alpha}^{\;\; \beta} = \lambda_{1}\;
\delta_{\alpha}^{\;\; \beta} + \lambda_{2} \;
F_{\alpha}^{\;\;\beta} + \lambda_{3} \; F_{\alpha}^{\;\;
\rho}F_{\rho}^{\;\; \beta}+ \lambda_{4} \; F_{\alpha}^{\;\;\rho}
F_{\rho}^{\;\; \sigma} F_{\sigma}^{\;\; \beta} \; . \label{4.6}
\end{eqnarray}

\noindent From the identity
\begin{eqnarray}
\Lambda_{\sigma}^{\;\; \alpha} \; (\Lambda^{-1})_{\alpha}^{\;\;
\beta} =
 \{ \mu  \delta_{\delta}^{\;\; \alpha} + \lambda F_{\sigma}^{\;\;
\alpha}\} \{\lambda_{1} \delta_{\alpha}^{\;\; \beta} + \lambda_{2}
F_{\alpha}^{\;\; \beta} + \lambda_{3} F_{\alpha}^{\;\;
\rho}F_{\rho}^{\beta}+ \lambda_{4} F_{\alpha}^{\; \;\rho}
F_{\rho}^{\;\; \chi} F_{\chi}^{\;\; \beta}\} \nonumber
\\
= \mu  \lambda_{1}\delta_{\sigma}^{\;\; \beta} + \mu \lambda_{2}
F_{\sigma}^{\;\; \beta} + \mu \lambda_{3} F_{\sigma}^{\;\; \rho}
F_{\rho}^{\;\; \beta}+ \mu \lambda_{4} F_{\sigma}^{\;\; \rho}
F_{\rho}^{\;\; \chi} F_{\chi}^{\;\; \beta} +
 \lambda \lambda_{1}\; F_{\sigma}^{\;\; \beta}
\nonumber
\\
+ \lambda \lambda_{2} \; F_{\sigma}^{\;\; \alpha}
F_{\alpha}^{\beta} + \lambda \lambda_{3} \; F_{\sigma}^{\;\;
\alpha} F_{\alpha}^{\;\; \rho} F_{\rho}^{\;\;   \beta}+
\lambda\lambda_{4}  \; F_{\sigma}^{\;\; \alpha} F_{\alpha}^{\;\;
\rho} F_{\rho}^{\;\; \chi} F_{\chi}^{\;\; \beta} =
\delta_{\sigma}^{\;\;\beta} \; . \label{4.7}
\end{eqnarray}

\noindent we get  the linear non-homogeneous system of equations
for  parameters  $\lambda_{1}, \lambda_{2}, \lambda_{3},
\lambda_{4}$:
\begin{eqnarray}
\mu \; \lambda_{1}+ \lambda \lambda_{4} \;  (\vec{E}\vec{B})^{2}=1
\; , \; \lambda \; \lambda_{1}+ \mu \; \lambda_{2}=0 \; ,
\nonumber
\\
\lambda\lambda_{2}+ \mu \lambda_{3}+\lambda\lambda_{4}
\;(\vec{E}^{2}-\vec{B}^{2})=0 \; , \; \lambda\lambda_{3}+ \mu
\; \lambda_{4}=0 \; . \label{4.8}
\end{eqnarray}

\noindent The  determinant of the main matrix is
\begin{eqnarray}
\hspace{30mm}
 \left | \begin{array}{rrrr}
\mu & 0 & 0 & \lambda(\vec{E}\vec{B})^{2} \\
\lambda & \mu & 0 & 0 \\
0 & \lambda& \mu & \lambda(\vec{E}^{2}-\vec{B}^{2}) \\
0 & 0 &  \lambda & \mu
 \end{array} \right |=  \mu^{2} \left [\mu^{2}-\lambda^{2}(\vec{E}^{2}-\vec{B}^{2})
\right ]-\lambda^{4}(\vec{E}\vec{B})^{2}\; . \label{4.9}
\end{eqnarray}

\noindent By Cramer's rule we obtain the solution of the linear
system:
\begin{eqnarray}
\lambda_{1}={ \mu[\mu^{2}-\lambda^{2}(\vec{E}^{2}-\vec{B}^{2})]
\over
 \mu^{2}[\mu^{2}-\lambda^{2}(\vec{E}^{2}-\vec{B}^{2})]-\lambda^{4}(\vec{E}\vec{B})^{2}} \; ,
\nonumber
\\
\lambda_{2}={-\lambda[\mu^{2}-\lambda^{2}(\vec{E}^{2}-\vec{B}^{2})]
\over  \mu^{2}[\mu^{2}-
\lambda^{2}(\vec{E}^{2}-\vec{B}^{2})]-\lambda^{4}(\vec{E}\vec{B})^{2}}
\; , \nonumber
\\
\lambda_{3}={\mu\lambda^{2} \over
\mu^{2}[\mu^{2}-\lambda^{2}(\vec{E}^{2}-\vec{B}^{2})]-
\lambda^{4}(\vec{E}\vec{B})^{2}} \; ,
\nonumber
\\
\lambda_{4}=-{\lambda^{3} \over
\mu^{2}[\mu^{2}-\lambda^{2}(\vec{E}^{2}-\vec{B}^{2})]-\lambda^{4}(\vec{E}\vec{B})^{2}}\;
. \label{4.10}
\end{eqnarray}

Additionally, we check the validity of the following relations:
\begin{eqnarray}
F_{\alpha}^{\;\;\beta}F_{\beta}^{\;\;\alpha}=
 2(\vec{E}^{2}-\vec{B}^{2})   \;,\qquad
 F_{\alpha}^{\;\; \beta}F_{\beta}^{\;\times
\rho}=4\vec{E}\vec{B}\delta_{\alpha}^{\beta}  \; , \qquad
\vec{E}\vec{B}={1 \over 4}
F_{\alpha}^{\;\;\beta}F_{\beta}^{\;\times\rho} \; . \nonumber
\end{eqnarray}

Thus, we arrive at the following explicit representation of the
inverse matrix:
\begin{eqnarray}
(\Lambda^{-1})_{\alpha}^{\;\;\beta}= {1\over \mu^{2}\left
(\mu^{2}-{\lambda^{2}\over
2}F_{\rho}^{\;\;\sigma}F_{\sigma}^{\;\;\rho} \right )-\lambda^{4}
\left ({1 \over 4}F_{\alpha}^{\;\;\beta}F_{\beta}^{\times \rho}
\right )^{2}}
\label{4.11}
\end{eqnarray}
$$
\times \left  \{
 \mu \left (\mu^{2}-{\lambda^{2} \over 2} F_{\rho}^{\;\;\sigma} F_{\sigma}^{\;\;\rho} \right )
\; \delta_{\alpha}^{\;\beta} - \lambda \left (
\mu^{2}-{\lambda^{2} \over 2} F_{\rho}^{\;\;\sigma}
F_{\sigma}^{\;\;\rho} \right )
 F_{\alpha}^{\;\;\beta} +
 \mu\lambda^{2} \;\; F_{\alpha}^{\;\;\sigma}F_{\sigma }^{\;\;\beta}
 -
 \lambda^{3} \;\; F_{\alpha}^{\;\;\sigma}F_{\sigma}^{\;\;\delta}F_{\delta}^{\;\;\beta} \right \} .
$$

Now let us consider the task of finding the inverse matrix
$ ( \Lambda ^{-1} )_{\alpha}^{\;\;\beta} $ in the case of Riemannian
space.   For simplicity, we will assume the metric tensor be
diagonal. It insures  the following property of the electromagnetic  tensor
$F_{\alpha \beta}$:
\begin{eqnarray}
F_{00}=0 \Longrightarrow F_{0}^{\;\;0}=0 \; , \qquad F_{11}=0
\Longrightarrow F_{1}^{\;\;1}=0 \; ,
\nonumber
\\
F_{22}=0
\Longrightarrow F_{2}^{\;\;2}=0 \; ,\qquad F_{33}=0
\Longrightarrow F_{3}^{\;\;3}=0 \; . \nonumber
\end{eqnarray}

\noindent We start from the explicit form of the tensor
$(F_{\alpha} ^{\;\; \beta})$:
\begin{eqnarray}
(F_{\alpha} ^{\;\; \beta}) = \left | \begin{array}{cccc}
0 & F_{0}^{\;\;1}  & F_{0}^{\;\;2}  & F_{0}^{\;\;3} \\
F_{1}^{\;\;0} & 0 & F_{1}^{\;\;2} & F_{1}^{\;\;3} \\
F_{2}^{\;\;0} & F_{2}^{\;\;1} & 0 &  F_{2}^{\;\;3} \\
F_{3}^{\;\;0} & F_{3}^{\;\;1} &  F_{3}^{\;\;2} & 0
 \end{array} \right | =
\left | \begin{array}{cccc}
 0 & g^{11}E_{1}  & g^{22}E_{2} & g^{33}E_{3} \\
-g^{00}E_{1}   & 0 & g^{22}B_{3}  & - g^{33} B_{2} \\
-g^{00}E_{2} & -g^{11} B_{3}  & 0 &  g^{33} B_{1}  \\
-g^{00}E_{3} & g^{11} B_{2} &  -g^{22} B_{1} & 0
 \end{array} \right | .
 \label{5.1}
 \end{eqnarray}

\noindent
 Below we  will use the notation
\begin{eqnarray}
g^{11}E_{1} = E^{1},  \; g^{22}E_{2} = E^{2}, \;
g^{33}E_{3} = E^{3}, \;\; g^{00}=h \; ,
\nonumber
\\
g^{22}g^{33} B_{1} = B^{1} \; , \;  g^{33}g^{11} B_{2} = B^{2} \;
, \; g^{11}g^{22} B_{3} = B^{3} \; .
\nonumber
\label{5.2}
\end{eqnarray}

Let us compute the convolution of two tensors (we write the matrix by the columns)
$$
(F_{\alpha} ^{\;\; \beta}) (F_{\beta} ^{\;\; \rho})=
\left | \begin{array}{rr}
-h E^{i}E_{i}  & -( E_{2}B^{3}- E_{3}B^{2})
 \\
-h(E^{2}B_{3}-E^{3}B_{2}) & -h E_{1} E^{1} -B_{2}B^{2}-B_{3}B^{3}
 \\
-h( E^{3}B_{1}-E^{1}B_{3}) &-h E^{1}E_{2}+B_{1}B^{2}
 \\
-h(E^{1}B_{2}-E^{2}B_{1}) & -h E^{1}E_{3}+B_{1}B^{3}
  \end{array} \right.;
$$
$$
\left. \begin{array}{rr}
 -(E_{3}B^{1} -E_{1}B^{3}) & -(E_{1}B^{2}-E_{2}B^{1}) \\
-hE_{1}E^{2}+B^{1}B_{2} & -hE_{1}E^{3}+B^{1}B_{3} \\
-hE_{2}E^{2}-B_{1}B^{1}-B_{3}B^{3}  & -h E_{2}E^{3}+B^{2}B_{3}\\
 -h E^{2}E_{3}+B_{2}B^{3} &  -hE_{3}E^{3}- B_{1} B^{1}-B_{2}B^{2}
  \end{array} \right |;
$$

\noindent next, compute the convolution of three tensors
\begin{eqnarray}
(F_{\alpha} ^{\;\; \beta}) (F_{\beta} ^{\;\; \rho})(F_{\rho}
^{\;\; \sigma}) =
 -( \; g^{00} E_{i}E^{i} + B_{i}B^{i} \; )\;
 \nonumber
 \\
 \times \left |
\begin{array}{rrrr}
0 & E^{1} & E^{2} &E^{3} \\
-g^{00}E_{1} & 0 & g^{22} B_{3} & -g^{33} B_{2} \\
-g^{00}E_{2}& -g^{11} B_{3} & 0 & g^{33} B_{1} \\
-g^{00}E_{3}& g^{11} B_{2} &  -g^{22} B_{1} & 0
 \end{array} \right |
 \nonumber
 \\
 + (B_{i}E_{i})\;
\left | \begin{array}{rrrr}
0 & g^{11} B^{1} & g^{22} B^{2} & g^{33} B^{3} \\
-g^{00}B^{1} & 0 & g^{22}g^{00} E^{3} & -g^{33}g^{00} E^{2} \\
-g^{00}B^{2}& -g^{11} g^{00} E^{3} & 0 & g^{33}g^{00}E^{1} \\
-g^{00}B^{3}& g^{11}g^{00} E^{2} &  -g^{22} g^{00} E^{1} & 0
 \end{array} \right |.
\label{5.4}
\end{eqnarray}

Using (\ref{5.3'}), we compute the convolution of two tensors in
two pairs of indices
\begin{eqnarray}
{1 \over 2} \; (F_{\alpha} ^{\;\;  \beta}F_{\beta} ^{\;\; \alpha}
)=- ( \; g^{00} E_{i}E^{i} + B_{i}B^{i} \; ) \equiv I(x)  \; .
\label{5.5}
\end{eqnarray}

Let us specify the dual  electromagnetic tensor
\begin{eqnarray}
(F^{\times})^{\alpha\beta} ={1\over 2} \epsilon^{\alpha\beta\rho
\sigma}(x)  F_{\rho \sigma} \; , \qquad \epsilon^{0123}(x)  =
\epsilon (x), \qquad \epsilon^{\alpha\beta\rho
\sigma}(x)=\epsilon^{[\alpha\beta\rho \sigma]}(x)\; , \label{5.6a}
\end{eqnarray}

\noindent where
\begin{eqnarray}
\epsilon (x) = {1 \over  \sqrt{-\det g }} = {1 \over\sqrt{-
g_{00}g_{11}g_{22}g_{33}}} = \sqrt{- g^{00}g^{11}g^{22}g^{33} }\;
. \label{5.6b}
\end{eqnarray}

\noindent Further,  we find explicit expressions for the
components of the dual tensor:
$$
(F^{\times})^{01}  =  \epsilon^{0123}(x) F_{23} =  \epsilon(x)
F_{23} =  \epsilon (x) B_{1}\; ,\qquad (F^{\times})_{0}^{\;\;1}  =
g_{00} \epsilon(x)      B_{1} = -\sqrt{-g}\;  g^{11} B^{1} \; ,
$$
$$
  (F^{\times})_{1}^{\;\;0}  =
- g_{11} \epsilon (x) B_{1} = + \sqrt{-g} \; g^{00}B^{1} \; ,
\qquad (F^{\times})^{02}  =  \epsilon^{0231}(x) F_{31} =
\epsilon(x) F_{31} =  \epsilon (x) B_{2}\; ,
$$
$$
 (F^{\times})_{0}^{\;\;2}  = g_{00} \epsilon(x)      B_{2} = -\sqrt{-g}\;g^{22} B^{2}\; ,
 \qquad
 (F^{\times})_{2}^{\;\;0} = - g_{22} \epsilon (x) B_{2} = + \sqrt{-g}\; g^{00}B^{2} \; ,
$$
$$
(F^{\times})^{03}  =  \epsilon^{0312}(x) F_{12} =  \epsilon(x)
F_{12} =  \epsilon (x) B_{3}\; , \qquad
 (F^{\times})_{0}^{\;\;3}  = g_{00} \epsilon(x)      B_{3} = - \sqrt{-g}\; g^{33} B^{3} \; ,
$$
$$
 (F^{\times})_{3}^{\;\;0} = - g_{33} \epsilon (x) B_{3} = +
\sqrt{-g}\; g^{00}B^{3} \; , \qquad
 (F^{\times})^{23}  =  \epsilon^{2301}(x) F_{01} =  \epsilon(x)
F_{01} =  \epsilon (x) E_{1}\; ,
$$
$$
(F^{\times})_{2}^{\;\;3} = g_{22} \epsilon (x) E_{1} =
-\sqrt{-g}\; g^{00}g^{33} E^{1} \; , \qquad
(F^{\times})_{3}^{\;\;2} = -g_{33} \epsilon (x) E_{1} = +
\sqrt{-g}\;g^{00}g^{22} E^{1} \; ,
$$
$$
(F^{\times})^{31}  =  \epsilon^{3102}(x) F_{02} =  \epsilon(x)
F_{02} =  \epsilon (x) E_{2}\; , \qquad (F^{\times})_{3}^{\;\; 1}
=    g_{33} \epsilon (x) E_{2} = -\sqrt{-g}\; g^{00}g^{11} E^{2}\;
,
$$
$$
 (F^{\times})_{1}^{\;\;
3} = -g_{11} \epsilon (x) E_{2} = + \sqrt{-g}\; g^{00}g^{33}
E^{2}\; , \qquad (F^{\times})^{12}  =  \epsilon^{1203}(x) F_{03} =
\epsilon(x) F_{03} =  \epsilon (x) E_{3}\; ,
$$
$$
(F^{\times})_{1}^{\;\;2}  = g_{11}  \epsilon (x) E_{3} =
-\sqrt{-g}\; g^{00}g^{22} E^{3}\;, \qquad
 (F^{\times})_{2}^{\;\;1}
= -g_{22}  \epsilon (x) E_{3} = +\sqrt{-g}\;  g^{00}g^{11} E^{3}\;
.
$$

So we obtain
$$
 {1 \over 4} (F_{\alpha} ^{\times  \beta}) (F_{\beta} ^{\;\; \alpha}) = {1 \over 4} \mbox{Sp}\left \{\;\; - \sqrt{-g}\;
\left | \begin{array}{cccc}
0 & g^{11} B^{1} & g^{22} B^{2} & g^{33} B^{3} \\
-g^{00}B^{1} & 0 & g^{00}g^{22} E^{3} & - g^{00}g^{33} E^{2} \\
-g^{00}B^{2} & - g^{00}g^{11} E^{3} & 0 & g^{00}g^{33} E^{1} \\
-g^{00}B^{3} & g^{00}g^{11} E^{2} & -g^{00}g^{22} E^{1} & 0
 \end{array} \right |  \right.
$$
$$
\left. \times \left | \begin{array}{rrrr}
 0 & E^{1}  & E^{2} & E^{3} \\
-g^{00} E_{1}   & 0 & g^{22}B_{3}  & - g^{33} B_{2} \\
-g^{00} E_{2} & -g^{11} B_{3}  & 0 &  g^{33} B_{1}  \\
-g^{00} E_{3} & g^{11} B_{2} &  -g^{22} B_{1} & 0
 \end{array} \right | \;\; \right \} \; ;
$$

\noindent let us consider  expressions for diagonal elements of
this product:
\begin{eqnarray}
(00)=  \sqrt{-g}\; g^{00}g^{11}g^{22}g^{33}  (E_{i}B_{i})= -{1
\over \sqrt{-g}}  (E_{i}B_{i} )\;, \qquad \qquad \nonumber
\\
 (11)= -{1 \over \sqrt{-g}} (E_{i}B_{i}) \;, \qquad  (22)=-{1 \over
\sqrt{-g}} (E_{i}B_{i}) \;, \qquad (33)=-{1 \over \sqrt{-g}}
(E_{i}B_{i}) \; . \nonumber
\end{eqnarray}

\noindent  Thus we arrive at the relationship
\begin{eqnarray}
 {1 \over 4} (F_{\alpha} ^{\times  \beta}) (F_{\beta} ^{\;\; \rho})=- {1 \over\sqrt{-  g }}\;(E_{i}B_{i})
 \equiv  J(x) \; .
\label{5.8}
\end{eqnarray}

Turning now to (\ref{5.4})
 we derive the following expansion for the convolution of
three tensors
\begin{eqnarray}
F_{\alpha} ^{\;\; \beta} F_{\beta} ^{\;\; \rho}F_{\rho} ^{\;\;
\sigma}=I(x)\; F_{\alpha}^{\;\;\sigma} + J(x) \;  F^{\times\;
\sigma}_{\alpha} \; ; \label{5.10}
\end{eqnarray}

\noindent where  the notation  $I(x), J(x)$ for the two invariants
of the electromagnetic field is used. An important point should be
stressed.   The representation (\ref{5.10}), due to the tensor
nature of the relation, is valid  in all (including non-orthogonal)
coordinate systems of space--time.

With the help of (\ref{5.10}) we  easily find the explicit form of
the convolution of the four tensors
\begin{eqnarray}
(F_{\alpha} ^{\;\; \beta} F_{\beta} ^{\;\; \chi}F_{\chi} ^{\;\;
\sigma}) (F_{\sigma} ^{\;\; \rho}) =
 I(x)\; F_{\alpha}^{\;\;\sigma} F_{\sigma} ^{\;\; \rho}
- J(x) \;  {1 \over \sqrt{-  g }}\;(E_{i}B_{i})
\delta_{\alpha}^{\;\;\rho} =
 I(x)\; F_{\alpha}^{\;\;\sigma} F_{\sigma} ^{\;\; \rho}
+ J^{2}(x) \;   \delta_{\alpha}^{\;\;\rho} \; . \label{5.11}
\end{eqnarray}

\noindent At the same time, the relation (\ref{5.11}) defines the
minimal polynomial
 of the matrix $(F_{\alpha}^{\;\;\beta})$.

The inverse of $ \Lambda_{\sigma}^{\;\;\alpha}= \mu
\delta_{\sigma}^{\;\;\alpha}+\lambda F_{\sigma}^{\;\;\alpha} $
tensor $(\Lambda^{-1})_{\alpha}^{\;\;\beta}$ should be constructed
in the form
\begin{eqnarray}
(\Lambda^{-1})_{\alpha}^{\;\; \beta} = \lambda_{1}\;
\delta_{\alpha}^{\;\; \beta} + \lambda_{2} \;
F_{\alpha}^{\;\;\beta} + \lambda_{3} \; F_{\alpha}^{\;\;
\rho}F_{\rho}^{\;\; \beta}+ \lambda_{4} \; F_{\alpha}^{\;\;\rho}
F_{\rho}^{\;\; \sigma} F_{\sigma}^{\;\; \beta} \; . \label{5.12}
\end{eqnarray}

\noindent From the identity
\begin{eqnarray}
\Lambda_{\sigma}^{\;\; \alpha} \; (\Lambda^{-1})_{\alpha}^{\;\;
\beta} = \{ \mu  \delta_{\delta}^{\;\; \alpha} + \lambda
F_{\sigma}^{\;\; \alpha}\} \{\lambda_{1} \delta_{\alpha}^{\;\;
\beta} + \lambda_{2} F_{\alpha}^{\;\; \beta} + \lambda_{3}
F_{\alpha}^{\;\; \rho}F_{\rho}^{\beta}+ \lambda_{4} F_{\alpha}^{\;
\;\rho} F_{\rho}^{\;\; \chi} F_{\chi}^{\;\; \beta}\} \nonumber
\\
= \mu  \lambda_{1}\delta_{\sigma}^{\;\; \beta} + \mu \lambda_{2}
F_{\sigma}^{\;\; \beta} + \mu \lambda_{3} F_{\sigma}^{\;\; \rho}
F_{\rho}^{\;\; \beta}+ \mu \lambda_{4} F_{\sigma}^{\;\; \rho}
F_{\rho}^{\;\; \chi} F_{\chi}^{\;\; \beta} \qquad\qquad \nonumber
\\
+ \lambda \lambda_{1}\; F_{\sigma}^{\;\; \beta} + \lambda
\lambda_{2} \; F_{\sigma}^{\;\; \alpha} F_{\alpha}^{\beta} +
\lambda \lambda_{3} \; F_{\sigma}^{\;\; \alpha} F_{\alpha}^{\;\;
\rho} F_{\rho}^{\;\;   \beta}+ \lambda\lambda_{4}  \;
F_{\sigma}^{\;\; \alpha} F_{\alpha}^{\;\; \rho} F_{\rho}^{\;\;
\chi} F_{\chi}^{\;\; \beta} = \delta_{\sigma}^{\;\;\beta} \; .
\label{5.13}
\end{eqnarray}

\noindent it follows  the linear non-homogeneous system of
equations for parameters $\lambda_{1}, \lambda_{2}, \lambda_{3},
\lambda_{4}$:
\begin{eqnarray}
\mu \lambda_{1}+ \lambda \lambda_{4} J^{2}{x}  =1 \; ,\;
\lambda\lambda_{1}+ \mu \lambda_{2}=0 \; , \;
\lambda\lambda_{2}+ \mu \lambda_{3}+\lambda\lambda_{4} \; I(x) =0
\; ,\; \lambda\lambda_{3}+ \mu\lambda_{4}=0  \; ; \label{5.14}
\end{eqnarray}

\noindent it  solution is
\begin{eqnarray}
\lambda_{1}={ \mu (\mu^{2}-\lambda^{2}  I  ) \over
 \mu^{2}( \mu^{2}-\lambda^{2} I )-\lambda^{4} J^{2} } \; , \qquad
\lambda_{2}={-\lambda ( \mu^{2}-\lambda^{2} I ) \over  \mu^{2}(
\mu^{2}-\lambda^{2} I )-\lambda^{4} J^{2} } \; , \nonumber
\\
\lambda_{3}={\mu\lambda^{2} \over \mu^{2}( \mu^{2}-\lambda^{2} I
)-\lambda^{4} J^{2} } \; , \qquad \lambda_{4}=-{\lambda^{3} \over
\mu^{2}( \mu^{2}-\lambda^{2} I )-\lambda^{4} J^{2} }\; .
\label{5.16}
\end{eqnarray}

Thus, we arrive at the following explicit representation of the
inverse tensor:
\begin{eqnarray}
(\Lambda^{-1})_{\alpha}^{\;\;\beta}= {1\over \mu^{2}  \; (\mu^{2}-
\lambda^{2} I  )-\lambda^{4}
 J^{2}} \qquad \qquad
 \nonumber
 \\
 \times \left  \{
 \mu  \; ( \; \mu^{2}- \lambda^{2} I  \; )
\; \delta_{\alpha}^{\;\beta} - \lambda  \; ( \mu^{2}- \lambda^{2}
I  )\;
 F_{\alpha}^{\;\;\beta} +
 \mu\lambda^{2} \; F_{\alpha}^{\;\;\sigma}F_{\sigma }^{\;\;\beta}
 -
 \lambda^{3} \;\; F_{\alpha}^{\;\;\sigma}F_{\sigma}^{\;\;\delta}F_{\delta}^{\;\;\beta} \right \} .
\label{5.17}
\end{eqnarray}

\noindent Using (\ref{5.10}),$\; F_{\alpha} ^{\;\; \sigma}  F_{\sigma}
^{\;\; \delta} F_{\delta}^{\;\; \beta}=
 I \;
F_{\alpha}^{\;\;\beta} + J \;  F^{\times\; \beta}_{\alpha} $, we
find a more simple representation
\begin{eqnarray}
(\Lambda^{-1})_{\alpha}^{\;\;\beta}= {1 \over
  \mu^{2}   (\mu^{2}- \lambda^{2} I  ) -\lambda^{4}
 J^{2} }
 \left \{
 \mu \;  (\;  \mu^{2}- \lambda^{2} I \;  )
\; \delta_{\alpha}^{\;\beta} - \lambda   \mu^{2}\;
 F_{\alpha}^{\;\;\beta} +
 \mu\lambda^{2}  \; F_{\alpha}^{\;\;\sigma}F_{\sigma }^{\;\;\beta}
 -
 \lambda^{3} \; J(x)  \;  F_{\alpha}^{\times  \beta} \right \}
 .
\nonumber
\\
\label{5.18}
\end{eqnarray}

These results   will be significantly simplified in the
case of a purely magnetic or purely electric field.

{\bf Magnetic field:}
\begin{eqnarray}
(\Lambda^{-1})_{\alpha}^{\;\;\beta}= {1 \over
  \mu^{2}   ( \mu^{2}- \lambda^{2} I  )   }  \left \{
 \mu \;  (\;  \mu^{2}- \lambda^{2} I \;  )
\; \delta_{\alpha}^{\;\beta} - \lambda   \mu^{2}\;
 F_{\alpha}^{\;\;\beta} +
 \mu\lambda^{2}  \; F_{\alpha}^{\;\;\sigma}F_{\sigma }^{\;\;\beta}
  \right \} ,
\nonumber
\\
 I(x)= -
( B_{i}B^{i} \; )   \; , \qquad
 J(x)= 0 \; ,\qquad
 (F_{\alpha} ^{\;\; \beta}) = \left | \begin{array}{rrrr}
 0 & 0  & 0 & 0 \\
0   & 0 & g^{22}B_{3}  & - g^{33} B_{2} \\
0 & -g^{11} B_{3}  & 0 &  g^{33} B_{1}  \\
0 & g^{11} B_{2} &  -g^{22} B_{1} & 0
 \end{array} \right | ,
\nonumber
\\
(F_{\alpha} ^{\;\; \sigma}) (F_{\sigma} ^{\;\; \beta} )= \left |
\begin{array}{rrrr}
0  &  0  & 0&  0\\
0 &  -B_{2}B^{2}-B_{3}B^{3} &
B^{1}B_{2}   & B^{1}B_{3} \\
0 &B_{1}B^{2}  &  -B_{1}B^{1}-B_{3}B^{3}
 & B^{2}B_{3} \\
0 & B_{1}B^{3} &  B_{2}B^{3} &  - B_{1} B^{1}-B_{2}B^{2}
 \end{array} \right |.
\label{5.19}
\end{eqnarray}

{\bf Electric field:}
\begin{eqnarray}
(\Lambda^{-1})_{\alpha}^{\;\;\beta}= {1 \over
  \mu^{2}   (\mu^{2}- \lambda^{2} I  )   }  \left \{
 \mu \;  (\;  \mu^{2}- \lambda^{2} I)
\; \delta_{\alpha}^{\;\beta} - \lambda   \mu^{2}\;
 F_{\alpha}^{\;\;\beta} +
 \mu\lambda^{2}  \; F_{\alpha}^{\;\;\sigma}F_{\sigma }^{\;\;\beta}
  \right \},
\nonumber
\\
 I= -
( \; g^{00} E_{i}E^{i})  \; ,\qquad
 J =  0
  \; ,\qquad
(F_{\alpha} ^{\;\; \beta}) = \left | \begin{array}{rrrr}
 0 & E^{1}  & E^{2} & E^{3} \\
-g^{00}E_{1}  & 0 & 0  & 0\\
-g^{00}E_{2} & 0 & 0 &  0 \\
-g^{00}E_{3} & 0 &  0 & 0
 \end{array} \right | ,
\nonumber
\\
 (F_{\alpha} ^{\;\; \sigma}) (F_{\sigma} ^{\;\; \beta} )=
g^{00} \left | \begin{array}{rrrr}
- E^{i}E_{i} & 0& 0& 0 \\
0 & - E_{1} E^{1} & -E_{1}E^{2} & -E_{1}E^{3}\\
0 &- E^{1}E_{2}  &  -E_{2}E^{2} & - E_{2}E^{3}\\
0  & - E^{1}E_{3}&  - E^{2}E_{3} &  -E_{3}E^{3}-
 \end{array} \right |.
\label{5.20}
\end{eqnarray}

Now we can construct in explicit form the generalized scalar equation, the tensor notation.
 Substituting the above tensor
$(\Lambda^{-1})_{\alpha } ^{\;\;\beta}$ into  the equation
(\ref{1.4b}) for a scalar function
  we obtain
\begin{eqnarray}
\left \{
 {\mu  \over \det \Lambda}
 \left ( \mu \;  (\;  \mu^{2}- \lambda^{2} I \;  ) \; \delta_{\alpha}^{\;\beta} -
\lambda   \mu^{2}\;  F_{\alpha}^{\;\;\beta} +
 \mu\lambda^{2}  \; F_{\alpha}^{\;\;\sigma}F_{\sigma }^{\;\;\beta}
 -
 \lambda^{3} \; J   \;  F_{\alpha}^{\times  \beta}
   \right )
D^{\alpha}  D_{\beta} + \right.
\label{6.2}
\end{eqnarray}
$$
\left. +    \left [ i \hbar \mu  \partial ^{\alpha} {1 \over \det
\Lambda}
 \left (\mu \;  (\;  \mu^{2}- \lambda^{2} I \;  ) \; \delta_{\alpha}^{\;\beta} -
\lambda   \mu^{2}\;  F_{\alpha}^{\;\;\beta} +
 \mu\lambda^{2}  \; F_{\alpha}^{\;\;\sigma}F_{\sigma }^{\;\;\beta}
 -
 \lambda^{3} \; J   \;  F_{\alpha}^{\times  \beta}  \right )
 \right  ] D_{\beta}
 - \mu^{2} \right \}   \Phi =0 \; .
$$

In the case of constant and uniform fields (in Cartesian
coordinates of the flat space), eq. (\ref{6.2}) becomes much more
simple:
\begin{eqnarray}
\left \{ \mu
 \left [
\mu \;  (\;  \mu^{2}- \lambda^{2} I \;  ) \;
\delta_{\alpha}^{\;\beta} - \lambda   \mu^{2}\;
F_{\alpha}^{\;\;\beta} +
 \mu\lambda^{2}  \; F_{\alpha}^{\;\;\sigma}F_{\sigma }^{\;\;\beta}
  -
 \lambda^{3} \; J   \;  F_{\alpha}^{\times  \beta}  \right ]
D^{\alpha}  D_{\beta}   - \mu^{2} \det \Lambda \right \}  \Phi =0
\; . \label{6.3}
\end{eqnarray}

The generalization of (\ref{6.2}) to the case of space-time models with
the Riemannian geometry is achieved by a formal change of ordinary
derivatives by covariant derivatives: $
\partial_{\alpha}\longrightarrow  \nabla_{\alpha} \;, \; D_{\alpha} = i \hbar \nabla_{\alpha} -
{e  \over c} A_{\alpha}\;, $ as a result we have
\begin{eqnarray}
 \mu
 \left [
 \mu \;  (\;  \mu^{2}- \lambda^{2} I \;  ) \; \delta_{\alpha}^{\;\beta} -
\lambda   \mu^{2}\;  F_{\alpha}^{\;\;\beta} +
 \mu\lambda^{2}  \; F_{\alpha}^{\;\;\sigma}F_{\sigma }^{\;\;\beta}
 -
 \lambda^{3} \; J   \;  F_{\alpha}^{\times  \beta}
   \right ]
D^{\alpha}  D_{\beta} \Phi \nonumber
\\
 +  i \hbar \mu  \; \det \Lambda \left [    \nabla _{\alpha}
 \left (
{\mu \;  (\;  \mu^{2}- \lambda^{2} I \over \det \Lambda } \;  ) \;
g^{\alpha \beta} - {\lambda   \mu^{2}\over \det \Lambda } \;
F^{\alpha \beta} \right. \right. \nonumber
\\
\left. \left.  -  {
 \lambda^{3} \; J \over \det \Lambda }   \;  F^{\times  \alpha\beta}
  +
 {\mu\lambda^{2} \over \det \Lambda }  \; g_{\rho \sigma}  F^{\alpha  \rho}F^{\sigma \beta}
    \right )
 \right  ] D_{\beta}
    \Phi  - \mu^{2}  \det \Lambda \;  \Phi \; =0 \; .
\label{6.4}
\end{eqnarray}

Let us specify the separate terms   under the symbol  of the
covariant derivative in
\begin{eqnarray}
 \left [    \nabla _{\alpha}
 \left (
{\mu \;  (\;  \mu^{2}- \lambda^{2} I \over \det \Lambda } \;  ) \;
g^{\alpha \beta} - {\lambda   \mu^{2}\over \det \Lambda } \;
F^{\alpha \beta}-{
 \lambda^{3} \; J \over \det \Lambda }   \;  F^{\times  \alpha\beta}
  +
 {\mu\lambda^{2} \over \det \Lambda }  \; g_{\rho \sigma}  F^{\alpha  \rho}F^{\sigma \beta}
    \right )
 \right  ] D_{\beta}
    \Phi  \; .
\nonumber
\end{eqnarray}

 The first term is reduced to the ordinary derivative of a scalar.
The second and third terms are the 4-divergences of the
antisymmetric tensor, and the fourth term is a 4-divergence of the
symmetric tensor; they are calculated using known formulas
\begin{eqnarray}
\nabla _{\alpha}
  ( A^{\alpha \beta }  ) =
( {1 \over \sqrt{-g} } \partial_{\alpha} \sqrt{-g} A^{\alpha
\beta} )  \; ,\qquad \nabla _{\alpha} ( S^{\alpha}_{\;\;  \beta} )
=
     ({1 \over \sqrt{-g}} \partial_{\alpha}  \sqrt{-g} S^{\alpha}_{\;\;  \beta}) -
{1 \over 2} ( \partial_{\beta} g_{\rho \sigma} ) S^{\rho \sigma}
  \; .
\label{6.5}
\end{eqnarray}

Now, let us derive non-relativistic  wave equation for Cox's particle, stating with
\begin{eqnarray}
  K_{\rho}^{\;\;\alpha} \left  (i\;  \nabla _{\alpha } -  {e\over c \hbar} A_{\alpha} \right )\; \Phi =
  { M c \over \hbar } \Phi_{\rho} \; ,
\qquad   \left (i\;  \nabla _{\alpha } -  {e\over c \hbar}
A_{\alpha} \right )\; \Phi^{\alpha}= {M c \over \hbar } \Phi \; .
\label{7.1}
\end{eqnarray}

\noindent The equations (\ref{7.1}) can be rewritten in a form
more convenient for practical calculations:
\begin{eqnarray}
 K_{\rho}^{\;\;\alpha} (i\;  \partial _{\alpha } -  {e\over c \hbar} A_{\alpha})
\Phi = {Mc \over  \hbar }    \Phi_{\rho} \; , \qquad ({i \over
\sqrt{-g}} \; {\partial \over  \partial x^{\alpha} } \sqrt{-g}  -
{e \over c \hbar } \; A_{\alpha})\;  g^{\alpha \beta} \Phi_{\beta}
= {Mc \over  \hbar }   \Phi \; . \label{7.2}
\end{eqnarray}

Considering the space-time models with the metric $ dS^{2} = c^{2} dt^{2}
+ g_{kl}(x)\;  dx^{k} dx^{l} $, in (\ref{7.2}) let us perform the splitting (3 +1)
\begin{eqnarray}
K_{0}^{\;\;0} (i\;  \partial  _{0 } -  {e\over c \hbar} A_{0})\;
\Phi  + K_{0}^{\;\;k} (i\;  \partial  _{k } -  {e\over c \hbar}
A_{k})\; \Phi
  =
{Mc\over \hbar }\; \Phi _{0 }  \; , \qquad \qquad \nonumber
\\
K_{j}^{\;\;0} (i\;  \partial  _{0 } -  {e\over c \hbar} A_{0})\;
\Phi  + K_{j}^{\;\;k} (i\;  \partial  _{k } -  {e\over c \hbar}
A_{k})\; \Phi
  =
{Mc\over \hbar }\; \Phi _{j }  \; , \qquad \qquad \nonumber
\\
({i \over \sqrt{-g}} \; {\partial \over  \partial x^{0} }
\sqrt{-g}  - {e \over c \hbar } \; A_{0})\;   \Phi_{0} +
 ({i \over \sqrt{-g}} \; {\partial \over  \partial x^{k} }
\sqrt{-g}  - {e \over c \hbar } \; A_{k})\;  g^{kl} \Phi_{l}= {Mc
\over  \hbar }   \Phi \; . \label{7.3}
\end{eqnarray}

\noindent Next, separate the rest energy by the
substitutions
\begin{eqnarray}
\Phi \Longrightarrow  \exp \; (- i {Mc^{2} t \over \hbar } ) \Phi
\; , \qquad \Phi_{0}   \Longrightarrow  \exp \; ( -i {Mc^{2} t
\over \hbar } ) \Phi_{0} \; , \qquad \Phi_{l} \Longrightarrow \exp
\; ( -i {Mc^{2} t \over \hbar } ) \Phi _{l}\; . \nonumber
\end{eqnarray}

\noindent As a result,  from  of (\ref{7.3}) we get
\begin{eqnarray}
K_{0}^{\;\;0} (i \hbar   \;  \partial  _{t } + Mc^{2}   -  e
A_{0})\; \Phi  + K_{0}^{\;\;k} (ic \hbar \;  \partial  _{k } -  e
A_{k})\; \Phi
  =
Mc^{2} \; \Phi _{0 }  \; , \qquad \qquad  \label{7.5a}
\\
K_{j}^{\;\;0} ( i\hbar  \;  \partial  _{t } +   Mc^{2}  -  e
A_{0})\; \Phi  + K_{j}^{\;\;k} (ic \hbar \;  \partial  _{k } -  e
A_{k})\; \Phi
  =
Mc^{2} \; \Phi _{j }  \; ,  \qquad \qquad \label{7.5b}
\\
\left  ( i \hbar  \partial_{t} +   Mc^{2}  + {i \hbar   \over
\sqrt{-g}} \; {\partial  \sqrt{-g}  \over  \partial t }
 - e  \; A_{0} \right )\;   \Phi_{0} +
  \left ({i c\hbar  \over \sqrt{-g}} \; {\partial \over  \partial x^{k} }
\sqrt{-g}  - e  \; A_{k} \right )\;  g^{kj } \Phi_{j}= Mc^{2}
   \Phi \; .
\label{7.5c}
\end{eqnarray}

\noindent Using (\ref{7.5b}), now  exclude the vector
(non-dynamic) variable $\Phi_{j}$:
\begin{eqnarray}
K_{0}^{\;\;0} (i \hbar   \;  \partial  _{t } + Mc^{2}   -  e
A_{0})\; \Phi  + K_{0}^{\;\;k} (ic \hbar \;  \partial  _{k } -  e
A_{k})\; \Phi
  =
Mc^{2} \; \Phi _{0 }  \; , \label{7.6a}
\\
\left ( i \hbar  \partial_{t} +   Mc^{2}  + {i \hbar   \over
\sqrt{-g}} \; {\partial \sqrt{-g} \over  \partial t }
  - e  \; A_{0} \right )\;   \Phi_{0} +
 \left ({i c\hbar  \over \sqrt{-g}} \; {\partial \over  \partial x^{k} }
\sqrt{-g}  - e  \; A_{k} \right ) \nonumber
\\
\times  {g^{kj} \over mc^{2} }
 \left [ K_{j}^{\;\;0} ( i\hbar  \;  \partial  _{t } +
Mc^{2}  -  e A_{0}) + K_{j}^{\;\;l} (ic \hbar \;  \partial  _{l }
-  e  A_{l}) \right ]\Phi = Mc^{2}    \Phi \; . \label{7.6b}
\end{eqnarray}

\noindent With  notation
\begin{eqnarray}
i \hbar   \;  \partial  _{t }    -  e  A_{0} = D_{t} \; , \qquad i
c \hbar   \;  \partial  _{k }    -  e  A_{k} = c  D_{k} \; ,
\qquad {i c\hbar  \over \sqrt{-g}} \; {\partial \over  \partial
x^{k} } \sqrt{-g}  - e  \; A_{k} = c \;\stackrel{\circ}{D}_{k} \;
, \label{7.7}
\end{eqnarray}

\noindent   equations (\ref{7.6a}), (\ref{7.6b}) can be
written as follows
\begin{eqnarray}
K_{0}^{\;\;0} (D_{t} + Mc^{2}  )\; \Phi  + K_{0}^{\;\;k} c D_{k}
\; \Phi
  =
Mc^{2}  \Phi _{0 }  \; , \qquad \qquad  \qquad
\nonumber
\label{7.8a}
\\
\left ( D_{t} + Mc^{2}    + {i \hbar   \over \sqrt{-g}} \;
{\partial \sqrt{-g} \over  \partial t }
 \right )   \Phi_{0} +
 \stackrel{\circ}{D}_{k} \;   {g^{kj} \over M }
\left ( {1 \over c} K_{j}^{\;\;0} ( D_{t} +   Mc^{2} ) +
K_{j}^{\;\;l} D_{l} \right ) \Phi = Mc^{2}    \Phi \; .
\label{7.8}
\end{eqnarray}

\noindent Following the method described in \cite{Red'kov}, we
introduce the small component  $\varphi$ and the big component
$\Psi$:
$
\Phi = (\Psi + \varphi )/2 , \; \Phi_{0} = (\Psi -
\varphi ) /2 $.
 Substituting these relations in
(\ref{7.8}):
$$
K_{0}^{\;\;0} D_{t}  {\Psi + \varphi   \over 2} +
 (K_{0}^{\;\;0} -1 +1) Mc^{2}  \; {\Psi + \varphi   \over 2}  +
 K_{0}^{\;\;k} c D_{k} \; {\Psi + \varphi   \over 2}
  =
Mc^{2} \; {\Psi - \varphi  \over 2}  \; ,
$$
$$
\left ( D_{t} + Mc^{2}    + {i \hbar   \over \sqrt{-g}} \;
{\partial  \sqrt{-g} \over  \partial t }
  \right )\;   {\Psi - \varphi  \over 2}
      +
 \stackrel{\circ}{D}_{k}    {g^{kj} \over M }
\left ( {1 \over c} K_{j}^{\;\;0} ( D_{t} +   Mc^{2} ) +
K_{j}^{\;\;l} D_{l} \right ) {\Psi + \varphi   \over 2} = Mc^{2}
{\Psi + \varphi  \over 2} \; .
$$

\noindent From whence, neglecting the small component $\varphi$ in
comparison with the  big $\Psi$, we obtain
\begin{eqnarray}
\left ( D_{t} + (K_{0}^{\;\;0} -1) (D_{t} +Mc^{2})   +
K_{0}^{\;\;k} c D_{k} \right )  {\Psi   \over 2}
  =-
Mc^{2} \;  \varphi \; , \label{7.11a}
\\
\left ( D_{t}   + {i \hbar   \over \sqrt{-g}} \; {\partial
\sqrt{-g} \over  \partial t }
  \right )\;   {\Psi   \over 2} +
 \stackrel{\circ}{D}_{k} \;   {g^{kj} \over M}
\left ( {1 \over c}  K_{j}^{\;\;0} ( D_{t} +   Mc^{2} ) +
K_{j}^{\;\;l} D_{l} \right ) {\Psi    \over 2} = Mc^{2}   \varphi
\; . \label{7.11b}
\end{eqnarray}

We assume that the energy of nonrelativistic particles are much
smaller than the rest energy,
 that is, we apply the approximation
$
 (D_{t} +Mc^{2})   \approx Mc^{2}
 $.
 As a result, the first and second equations are simplified
\begin{eqnarray}
\left ( D_{t} + (K_{0}^{\;\;0} -1) Mc^{2}   + K_{0}^{\;\;k} c
D_{k} \right )  {\Psi   \over 2}
  =-
Mc^{2} \;  \varphi \; , \label{7.12a}
\\
\left ( D_{t}   + {i \hbar   \over \sqrt{-g}} \; {\partial
\sqrt{-g} \over  \partial t } \right  )   {\Psi   \over 2} +
 \stackrel{\circ}{D}_{k} \;   {g^{kj} \over M}
\left (   M c  K_{j}^{\;\;0}     + K_{j}^{\;\;l} D_{l} \right )
{\Psi    \over 2} = Mc^{2}   \varphi   \; . \label{7.12b}
\end{eqnarray}

\noindent With the help of  (\ref{7.12a}), one  can eliminate the
small component $\varphi$ from the equation (\ref{7.12b}), so we
get
\begin{eqnarray}
( D_{t}   + {i \hbar   \over \sqrt{-g}} \; {\partial \over
\partial t } \sqrt{-g}  )\;   {\Psi   \over 2} +
 \stackrel{\circ}{D}_{k}    {g^{kj} \over M}
\left ( Mc K_{j}^{\;\;0}  + K_{j}^{\;\;k} D_{k} \right ) {\Psi
\over 2} =
 - \left ( D_{t} + (K_{0}^{\;\;0} -1) Mc^{2}   + K_{0}^{\;\;k} c
D_{k} \right )  {\Psi   \over 2}   \; ,
\nonumber
 \label{7.13}
\end{eqnarray}

\noindent or
\begin{eqnarray}
\left ( D_{t}   + {i \hbar   \over 2 \sqrt{-g}} \; {\partial
\sqrt{-g} \over  \partial t }
  \; + {1 \over 2} [ (K_{0}^{\;\;0} -1)  Mc^{2}   +
 K_{0}^{\;\;k} c D_{k}  ] \right ) \Psi  =
 \stackrel{\circ}{D}_{k} \;   { (-g^{kj}) \over 2M }
\left [ K_{j}^{\;\;l} D_{l}  + Mc  K_{j}^{\;\;0} \right ] \Psi \;
. \qquad \label{7.14}
\end{eqnarray}

\noindent With the substitution $ \Psi  \; \Longrightarrow \;
(-g)^{-1/4} \; \Psi $, equation (\ref{7.14}) reads simpler
\begin{eqnarray}
 D_{t}    \;  \Psi ={1 \over 2M}
 \stackrel{\circ}{D}_{k}     (-g^{kj})
\left ( K_{j}^{\;\;l} D_{l}  + Mc  K_{j}^{\;\;0} \right ) \Psi
- {1 \over 2} \left (   (K_{0}^{\;\;0} -1)  Mc^{2}   +
 K_{0}^{\;\;j} c D_{j}   \right ) \Psi \; .
\label{7.15}
\end{eqnarray}
This is the nonrelativistic Schr\"{o}dinger equation for  the
Cox's particle.

Let us specify the  case of the presence of only
 magnetic field, then  the Schr\"{o}dinger
equation (\ref{7.15}) becomes  much simpler
\begin{eqnarray}
 D_{t}    \;  \Psi ={1 \over 2M}
 \stackrel{\circ}{D}_{k}     (-g^{kj}) K_{j}^{\;\;l} D_{l} \;  \Psi \; .
\label{7.16}
\end{eqnarray}

\noindent Let us detail the operator $ K_{j}^{\;\;l} D_{l}$:
\begin{eqnarray}
K_{1}^{\;\;l} D_{l} =  K_{1}^{\;\;1} D_{1}+ K_{1}^{\;\;2} D_{2} +
K_{1}^{\;\;3} D_{3} =
 {1 \over \mu^{2} +  \lambda^{2} B_{i} B^{i} } [ \; \mu^{2} D_{1}
+ \mu \lambda (B_{2} D^{3} - B_{3} D^{2}) +   \lambda^{2} B^{1} \;
(B_{i} D_{i} )\; ] \; , \nonumber
\\
K_{2}^{\;\;l} D_{l} =  K_{2}^{\;\;1} D_{1}+ K_{2}^{\;\;2} D_{2} +
K_{2}^{\;\;3} D_{3} = {1 \over \mu^{2} +  \lambda^{2} B_{i}B^{i}
}[ \; \mu^{2} D_{2} + \mu \lambda (B_{3} D^{1} - B_{1} D^{3}) +
\lambda^{2} B^{2} \; (\vec{B} \vec{D} )\; ]  \; , \nonumber
\\
K_{3}^{\;\;l} D_{l} =  K_{3}^{\;\;1} D_{1}+ K_{3}^{\;\;2} D_{2} +
K_{3}^{\;\;3} D_{3} =
 {1 \over \mu^{2} +  \lambda^{2} B_{i}B^{i}} [\; \mu^{2} D_{3} +
\mu \lambda (B_{1} D^{2} - B_{2} D^{1}) +   \lambda^{2} B^{3} \;
(B_{i} D_{i} ) \; ] \; . \nonumber
\end{eqnarray}

\noindent Thus, we have (we use the notation $\Gamma = \lambda /
\mu$)
\begin{eqnarray}
\stackrel{\ast}{D}_{1}= K_{1}^{\;\;l} D_{l} ={1 \over 1  +
\Gamma^{2} B_{i} B^{i} } \left  [ \; D_{1} + \Gamma (B_{2} D^{3} -
B_{3} D^{2}) +   \Gamma^{2} B^{1} \; (B_{i} D_{i} )\; \right ] \;
, \nonumber
\\
\stackrel{\ast}{D}_{2} = K_{2}^{\;\;l} D_{l} = {1 \over 1  +
\Gamma^{2} B_{i}B^{i} } \left [ \; D_{2} +  \Gamma (B_{3} D^{1} -
B_{1} D^{3}) +   \Gamma^{2} B^{2} \; (\vec{B} \vec{D} )\;\right  ]
\; , \nonumber
\\
\stackrel{\ast}{D}_{3} = K_{3}^{\;\;l} D_{l} = {1 \over 1 +
\Gamma^{2} B_{i}B^{i}} \left [\;
 D_{3} + \Gamma (B_{1} D^{2} - B_{2} D^{1}) +   \Gamma^{2} B^{3} \; (B_{i} D_{i} ) \;\right  ] \; .
\label{7.17}
\end{eqnarray}

\noindent Therefore, the equation (\ref{7.16}) can be written in a
compact form as follows
\begin{eqnarray}
 D_{t}    \;  \Psi = - {1 \over 2 M}
 \stackrel{\circ}{D}_{k}     g^{kj}(x) \stackrel{\ast}{D}_{j}    \Psi \; .
\label{7.18}
\end{eqnarray}

In the case of the Cartesian coordinates of the flat space, metric
tensor is trivial, and the equation is simplified
\begin{eqnarray}
 D_{t}    \;  \Psi =  {1 \over 2 M}
 \vec{D}  \left ({  \vec{D}
 -   \Gamma \;  \vec{B} \times  \vec{D}  +\Gamma^{2}   \vec{B}  (\vec{B} \vec{D} )\over 1
  +  \Gamma^{2} \vec{B}^{\;2}} \right )   \Psi \; .
\label{7.19}
\end{eqnarray}

\noindent
If the magnetic field in the flat space is uniform, equation
(\ref{7.19}) can be simplified yet  more
\begin{eqnarray}
  D_{t}    \;  \Psi =  {1 \over 2M} {1 \over 1 +  \Gamma^{2} \vec{B}^{\;2}}
   \left (  \vec{D}^{2}
 -   \Gamma \;  \vec{D} (\vec{B} \times  \vec{D})  +\Gamma^{2} \;    (\vec{B} \vec{D})^{2} \right )   \Psi \; .
\label{7.20}
\end{eqnarray}

\noindent Due to the identity $ \vec{D}  (\vec{B}\times  \vec{D})=
+i{e \hbar \over c} \vec{B}^{\;2}\;, $ the equation (\ref{7.20})
can be represented as follows:
\begin{eqnarray}
  D_{t}    \;  \Psi =  {1 \over 2M} {1 \over 1 +  \Gamma^{2} \vec{B}^{\;2}}
   \left (  \vec{D}^{2}
 -  i{e \hbar \over c}  \Gamma \;  \vec{B}^{\;2}  +\Gamma^{2} \;    (\vec{B} \vec{D})^{2} \right )   \Psi \; .
\label{7.22}
\end{eqnarray}

Note that the presence of the term $
 i (e \hbar /c)  \Gamma \;  \vec{B}^{\;2}
 $  means that the parameter $\Gamma$ must  be purely imaginary.
  The explicit form of (\ref{7.22}) also implies that there is a
  steady  shift of all levels on the  value determined by the amplitude of the magnetic
  field and the parameter $i\Gamma$.

Now consider the case of a uniform electric field.The operator $
K_{j}^{\;\;l} D_{l}  + \mu  K_{j}^{\;\;0}  $ is
\begin{eqnarray}
K_{1}^{\;\;l} D_{l}  + \mu  K_{1}^{\;\;0} ={ 1 \over  1 + \Gamma^{2} E_{i}E^{i} }   \left [\; D_{1} +
\Gamma^{2}  (E_{i}E^{i}) D_{1} + \Gamma^{2}  E_{1} (E^{i} D_{i}) +
 \mu   \Gamma E_{1}) \;\right  ]    \; ,
\nonumber
\\
K_{2}^{\;\;l} D_{l}  + \mu  K_{2}^{\;\;0} = {1 \over 1 +
\Gamma^{2} E_{i}E^{i} } \left [\; D_{2} +    \Gamma^{2}
(E_{i}E^{i}) D_{2} + \Gamma^{2}  E_{2} (E^{i} D_{i})   +
 \mu   \Gamma E_{2}) \;\right  ] \; ,
\nonumber
\\
K_{3}^{\;\;l} D_{l}  + \mu  K_{3}^{\;\;0} = {1 \over 1 +
\Gamma^{2} E_{i}E^{i} } \left [\; D_{3} +    \Gamma^{2}
(E_{i}E^{i}) D_{3} + \Gamma^{2}  E_{3} (E^{i} D_{i})   +
 \mu   \Gamma E_{3}) \;\right  ] \; ,
\nonumber
\end{eqnarray}

\noindent thus
\begin{eqnarray}
( K_{j}^{\;\;l} D_{l}   + \mu  K_{j}^{\;\;0}) = {1 \over 1 +
\Gamma^{2} E_{i}E^{i} } \left [\; D_{j} +    \Gamma^{2}
(E_{i}E^{i}) D_{j} + \Gamma^{2}  E_{j} (E^{i} D_{i})   +
 \mu   \Gamma E_{j} \;\right  ] .
\label{7.24}
\end{eqnarray}

\noindent So we have the representation
\begin{eqnarray}
   (K_{0}^{\;\;0} -1)  Mc^{2}   +
 K_{0}^{\;\;j} c D_{j}   = -c\;
 {
  \Gamma ^{2} E_{i}E^{i}  \mu  +
     \Gamma    E^{j} D_{j}  \over 1 +  \Gamma^{2} E_{i}E^{i} }   \;  .
 \label{7.25}
 \end{eqnarray}

Thus, we get
\begin{eqnarray}
\left (  D_{t}  - c\;
 {   \Gamma ^{2} E_{i}E^{i}  \mu  +
     \Gamma    E^{j} D_{j}  \over 2( 1 +  \Gamma^{2} E_{i}E^{i}  )}  \right )     \Psi
          = {1 \over 2M}
 \stackrel{\circ}{D}_{k}     (-g^{kj})
\left [\; D_{j}  + { \Gamma^{2}  E_{j} (E^{i} D_{i})   +
 \mu   \Gamma E_{j}  \over 1 + \Gamma^{2} E_{i}E^{i} }   \;\right  ] \Psi\, ;
\label{7.26}
\end{eqnarray}
this is the  Schr\"{o}dinger equation for a Cox's particle  in the
electric field.

No we examine the initial complete Cox's system of  equations \cite{Cox-1982}
  which includes  symmetric and antisymmetric tensors:
 \begin{eqnarray}
 \lambda_{1} D^{\beta} \Phi_{\beta} - \mu  \Phi =0  ,\qquad
 \lambda_{1}^{*} D_{\beta} \Phi + \lambda_{2} D^{\alpha} \Phi_{[\alpha \beta]} -
 \lambda_{3} D^{\alpha}   \Phi_{(\alpha \beta)} - \mu  \Phi_{\beta}=0, \qquad\quad
\nonumber
\\
\lambda_{2}^{*} ( D_{\alpha} \Phi_{\beta} - D_{\beta}
\Phi_{\alpha} ) - \mu  \Phi _{[\alpha \beta]}=0, \qquad
\lambda_{3}^{*} ( D_{\alpha} \Phi_{\beta} + D_{\beta}
\Phi_{\alpha} -{1 \over 2}g_{\alpha \beta} D^{\rho} \Phi_{\rho}) -
  \mu  \Phi _{(\alpha \beta)}=0,
\label{8.1}
\end{eqnarray}

\noindent where the auxiliary numerical parameters $\lambda_ {1},
\lambda_{2}, \lambda_ {3}$
 subject to the additional constraints
\begin{eqnarray}
\lambda_{2}\lambda_{2}^{*} - \lambda_{3} \lambda_{3}^{*}  =0 \;
,\; \lambda_{1}\lambda_{1}^{*} - {3 \over 2} \lambda_{3}
\lambda_{3}^{*} = 1 \; ; \label{8.2}
\end{eqnarray}

\noindent symbol $D_ {\alpha}$ denotes the derivative, which takes
into account the presence of external electromagnetic and
gravitational fields
\begin{eqnarray}
D_{\alpha}  =i\hbar \nabla_{\alpha} - { e\over c} A_{\alpha} \; ,
\qquad \mu = Mc \; . \nonumber
\end{eqnarray}

With the help of the third and the fourth equations in (\ref{8.1}), let
us  exclude tensor components
\begin{eqnarray}
\mu^{-1} \left ( \lambda_{2}^{*} ( D_{\alpha} \Phi_{\beta} -
D_{\beta} \Phi_{\alpha} )  \right )=  \Phi _{[\alpha \beta]},
\nonumber
\\
\mu^{-1}  \lambda_{3}^{*} ( D_{\alpha} \Phi_{\beta} + D_{\beta}
\Phi_{\alpha} -{1 \over 2}g_{\alpha \beta} D^{\rho} \Phi_{\rho})
=   \Phi _{(\alpha \beta)},
 \label{8.3}
 \end{eqnarray}

\noindent in  the two other:
\begin{eqnarray}
 \lambda_{1} D^{\beta} \Phi_{\beta} - \mu  \Phi =0 \; ,
  \label{8.4}
  \end{eqnarray}
  \begin{eqnarray}
    \lambda_{1}^{*} D_{\beta} \Phi + \lambda_{2} D^{\alpha}  \mu^{-1} \left [
  \lambda_{2}^{*} ( D_{\alpha} \Phi_{\beta} - D_{\beta} \Phi_{\alpha} )  \right ]
  \nonumber
  \\
  -
 \lambda_{3} D^{\alpha}   \mu^{-1}  \lambda_{3}^{*}
  ( D_{\alpha} \Phi_{\beta} + D_{\beta} \Phi_{\alpha}
-{1 \over 2}g_{\alpha \beta} D^{\rho} \Phi_{\rho})
 - \mu
\Phi_{\beta}=0\; .
  \label{8.5}
  \end{eqnarray}
  Eq. (\ref{8.5}) can be presented as
\begin{eqnarray}
 \lambda_{1}^{*} D_{\beta} \Phi - \mu^{-1} (\lambda_{2}  \lambda_{2}^{*} + \lambda_{3} \lambda_{3}^{*} ) \;
    D^{\alpha} D_{\beta} \Phi_{\alpha}
+ {1 \over 2} \mu^{-1}  \lambda_{3} \lambda_{3}^{*} \;     \;
D_{\beta}  D^{\rho} \Phi_{\rho}  - \mu  \Phi_{\beta}=0\,.
  \label{8.6a}
  \end{eqnarray}

\noindent In view of (\ref{8.2}), one  can apply identity
\begin{eqnarray}
(\lambda_{2}  \lambda_{2}^{*} + \lambda_{3} \lambda_{3}^{*} )= 2
\lambda_{3} \lambda_{3}^{*} \; ; \nonumber
\end{eqnarray}

\noindent hence
\begin{eqnarray}
 \lambda_{1}^{*} D_{\beta} \Phi - \mu^{-1} 2 \lambda_{3} \lambda_{3}^{*}  \;
    D_{\alpha} D_{\beta} \Phi^{\alpha}
        + {1 \over 2} \mu^{-1}  \lambda_{3} \lambda_{3}^{*} \;
D_{\beta}  D_{\alpha} \Phi^{\alpha}  - \mu  \Phi_{\beta}=0\,.
  \label{8.6b}
  \end{eqnarray}

We use the identity
\begin{eqnarray}
D_{\alpha} D_{\beta} \Phi^{\alpha} = D_{\beta} D_{\alpha}
\Phi^{\alpha}
 +  ( D_{\alpha} D_{\beta}  - D_{\beta} D_{\alpha}  ) \Phi^{\alpha}
  =
 D_{\beta} D_{\alpha}   \Phi^{\alpha} + \hbar ^{2}  \left ( -i  {e \over \hbar c} F_{\alpha \beta} - R_{\alpha \beta} \right )
 \Phi^{a\alpha}  ,
\nonumber
\\
\label{8.7}
\end{eqnarray}

\noindent equation (\ref{8.6b}) can be converted to the following
one
\begin{eqnarray}
  \lambda_{1}^{*} D_{\beta} \Phi +\mu^{-1} 2 \lambda_{3} \lambda_{3}^{*}  \;
    \hbar ^{2}  \left (  i  {e \over c} F_{\alpha \beta} + R_{\alpha \beta} \right )
 \Phi^{\alpha}
  - {3 \over 2} \mu^{-1}  \lambda_{3} \lambda_{3}^{*} \;
D_{\beta}  (D_{\alpha} \Phi^{\alpha})
  - \mu  \Phi_{\beta}=0\; .\quad
\label{8.8}
\end{eqnarray}

Taking  into account the equation (\ref{8.4})
\begin{eqnarray}
  D_{\alpha} \Phi^{\alpha} = {\mu   \over \lambda_{1}} \Phi  \; ,
 \nonumber
 \end{eqnarray}

\noindent one produces
\begin{eqnarray}
  \lambda_{1}  \lambda_{1}^{*} D_{\beta} \Phi
   + \mu^{-1} 2 \lambda_{3} \lambda_{3}^{*}  \;
    \hbar ^{2}  \left (i  {e \over \hbar c} F_{\alpha \beta} + R_{\alpha \beta} \right )
 \lambda_{1} \Phi^{\alpha}
 - {3 \over 2}  \lambda_{3} \lambda_{3}^{*} \;     D_{\beta}  \Phi
  - \mu  \lambda_{1} \Phi_{\beta}=0\; .
\label{8.9a}
\end{eqnarray}

\noindent With the use of the second condition in (\ref{8.2})
\begin{eqnarray}
\lambda_{1}\lambda_{1}^{*} - {3 \over 2} \lambda_{3}
\lambda_{3}^{*} = 1 \nonumber
\end{eqnarray}

\noindent on simplifies eq.  (\ref{8.9a})  to the form
\begin{eqnarray}
  D_{\beta} \Phi +  \mu^{-1} 2 \lambda_{3} \lambda_{3}^{*}  \;
    \hbar ^{2}  \left ( i  {e \over \hbar c} F_{\alpha \beta} + R_{\alpha \beta} \right )
 \lambda_{1} \Phi^{\alpha}
   - \mu  \lambda_{1} \Phi_{\beta}=0\; . \qquad \qquad
\label{8.9b}
\end{eqnarray}

\noindent One should remember on  additional equation (\ref{8.4})
\begin{eqnarray}
 \lambda_{1} D^{\beta} \Phi_{\beta} - \mu  \Phi =0 \; .
  \label{8.9c}
  \end{eqnarray}

The parameter $\lambda_ {1}$ can be included in the designation of
vector components
$
\lambda_{1} \Phi_{\beta} \longrightarrow \Phi_{\beta} \,;$
 so we arrive at the extended Proca equations
\begin{eqnarray}
 D^{\beta} \Phi_{\beta} - \mu  \Phi =0 \,,\quad
  D_{\beta} \Phi  - \mu   \Phi_{\beta}
   -i   {   \hbar^{2}
       \over Mc }  (2 \lambda_{3} \lambda_{3}^{*})
 \left (    {e \over  \hbar c} F_{\beta \alpha } + i R_{\beta \alpha } \right )
  \Phi^{\alpha}
    =0\; .
\label{8.10}
\end{eqnarray}

\noindent
These equations should be compared with Proca-like system used above
\begin{eqnarray}
 D^{\beta} \Phi_{\beta} - \mu  \Phi =0 \; ,
 \quad
  D_{\beta} \Phi - \mu   \Phi_{\beta} - \lambda   F_{\beta \alpha }   \Phi^{\alpha}
  =0\; ;
\label{8.10'}
  \end{eqnarray}
they  correlate if (note that $\lambda$ is an imaginary number)
\begin{eqnarray}
  \lambda =   {   \hbar^{2}       \over Mc }   {e \over  \hbar c}  (2i \lambda_{3} \lambda_{3}^{*}) \; .
\label{8.11}
\end{eqnarray}

Obviously, the system (\ref{8.10}) is more general than (\ref{8.10'}),  it takes into
 account non-minimal interaction of the  Cox's scalar particle with external geometric
  background through the Ricci tensor.

Equations (\ref{8.10}) can be rewritten as
\begin{eqnarray}
 D^{\beta} \Phi_{\beta} - \mu  \Phi =0 \,, \quad
  D_{\beta} \Phi - \lambda  \left (    F_{\beta \alpha } + i {\hbar c \over e} R_{\beta \alpha } \right )
  \Phi^{\alpha}
  - \mu   \Phi_{\beta}=0\; .
\label{8.12}
\end{eqnarray}

In the absence of the electromagnetic field,
 equations (\ref{8.12}) are simplified (parameter $i\lambda$ is a real-valued)
\begin{eqnarray}
 D^{\beta} \Phi_{\beta} =  \mu  \Phi  \; , \quad
   D_{\beta} \Phi  = \left  (  i\lambda    {\hbar c \over e} R_{\beta \alpha }(x)
  + \mu \;  g_{\beta \alpha} (x)   \right )   \Phi^{\alpha} \; .
\label{8.13}
\end{eqnarray}
This is purely geometric modification of the theory of a scalar
particle in the Cox's approach.

Finally, let us find the  tensor
$(\Lambda ^{-1})_{\alpha}^{\;\;\beta}$ when the  Ricci tensor
is taken into account. To this end, we write equation (\ref{8.13}) in the form ($\lambda^{*}
 = - \lambda$; temporarily the  coefficient ${\hbar c \over e}$ will be a part of designation of the
  Ricci tensor)
\begin{eqnarray}
 D^{\beta} \Phi_{\beta} = \mu  \Phi  \;,
\quad
 \; [ \mu  \delta_{\alpha}^{\; \beta} + \lambda (F_{\alpha}^{\;\; \beta} + i R_{\alpha}^{\;\;\beta})] \Phi_{\beta}
 = D_{\alpha} \Phi  \; .
\label{9.1}
\end{eqnarray}

\noindent
With the use of the notation
 \begin{eqnarray}
 \Lambda_{\alpha}^{\;\;\beta} = \mu  \delta_{\alpha}^{\; \beta} +
 \lambda (F_{\alpha}^{\;\; \beta} + i R_{\alpha}^{\;\;\beta})
 \label{9.2a}
 \end{eqnarray}
  equation (\ref{9.1}) can be written as:
\begin{eqnarray}
\Phi_{\rho} = (\Lambda^{-1})_{\rho}^{\;\;\alpha}D_{\alpha}\Phi \;
, \qquad D^{\rho} \Phi_{\rho}=\mu \Phi \; . \label{9.2b}
\end{eqnarray}

\noindent From this it  follows a generalized scalar equation
\begin{eqnarray}
\left [  D^{\rho}(\Lambda^{-1})_{\rho}^{\;\; \alpha}(x) D_{\alpha}
\; - \; \mu \right ]  \Phi (x) =0\; . \label{9.3}
\end{eqnarray}

Since the characteristic equation for the matrix $F_{\alpha}^{\;\; \beta} + i R_{\alpha}^{\;\;\beta} = G_{\alpha
}^{\;\;\beta} $:
\begin{eqnarray}
G^{4} =  g_{0}  + g_{1} G + g_{2} G^{2} + g_{3} G^{3} \label{9.4a}
\end{eqnarray}
allows us to express the fourth power of the matrix $G$ through
$I, G, G^{2}, G^{3}$, we can look for the inverse matrix in the
form
\begin{eqnarray}
(\Lambda^{-1})_{\rho}^{\;\;\alpha} = \lambda_{0} + \lambda_{1} G +
\lambda_{2} G^{2} + \lambda_{3} G^{3} \; . \label{9.4b}
\end{eqnarray}

\noindent From the equation $\Lambda \Lambda ^{-1} =I$ if follows
\begin{eqnarray}
I = (\mu + \lambda G) (\lambda_{0} + \lambda_{1} G + \lambda_{2}
G^{2} + \lambda_{3} G^{3} ) \nonumber
\\
= \mu  \lambda_{0} +  \mu  \lambda_{1} G +  \mu  \lambda_{2} G^{2}
+ \mu   \lambda_{3} G^{3} \nonumber
\\
+ \lambda \lambda_{0} G  + \lambda \lambda_{1} G^{2}  +  \lambda
\lambda_{2} G^{3} \nonumber
\\
+ \lambda \lambda_{3} (g_{0}  + g_{1} G + g_{2} G^{2} + g_{3}
G^{3})\; ; \nonumber
\end{eqnarray}

\noindent so  we obtain the linear system
\begin{eqnarray}
I : \qquad \qquad  \mu \lambda_{0} + \lambda \lambda_{3} g_{0}  =
1 \; , \nonumber
\\
G: \qquad  \mu \lambda_{1} +\lambda \lambda_{0} + \lambda
\lambda_{3} g_{1} =0  \; , \nonumber
\\
G^{2}: \qquad  \mu \lambda_{2} + \lambda \lambda_{1} + \lambda
\lambda_{3} g_{2} =0  \; , \nonumber
\\
G^{3}: \qquad  \mu \lambda_{3} + \lambda \lambda_{2} + \lambda
\lambda_{3} g_{3} =0  \;  . \label{9.5a}
\end{eqnarray}

We write the system in the  matrix form
\begin{eqnarray}
\left ( \begin{array}{cccc}
\mu  & 0 & 0& \lambda g_{0}  \\
\lambda  & \mu  & 0& \lambda g_{1} \\
0 & \lambda  & \mu & \lambda g_{2} \\
0 & 0 & \lambda & \mu + \lambda g_{3}
\end{array} \right )
 \left | \begin{array}{c}
\lambda_{0}\\ \lambda_{1}\\ \lambda_{2}\\ \lambda_{3}
\end{array} \right | = \left | \begin{array}{c} 1 \\ 0 \\ 0 \\ 0
\end{array} \right |.
\label{9.5b}
\end{eqnarray}

\noindent Its solution is
\begin{eqnarray}
\lambda_{0} = { - (\mu^{3}+\mu^{2} \lambda g_{3}-\mu \lambda^{2}
g_{2}+\lambda^{3} g_{1}) \over - \mu^{4}-\mu^{3} \lambda
g_{3}+\mu^{2} \lambda^{2} g_{2}-\mu \lambda^{3}
g_{1}+\lambda^{4}g_{0}}\,, \nonumber
\\
\lambda_{1} = {-( -\mu^{2} \lambda-\mu \lambda^{2}
g_{3}+\lambda^{3} g_{2} )\over - \mu^{4}-\mu^{3} \lambda
g_{3}+\mu^{2}  \lambda^{2} g_{2}-\mu \lambda^{3} g_{1}+\lambda^{4}
g_{0}}\,, \nonumber
\\
\lambda_{2} = {-(\mu \lambda^{2}+\lambda^{3}g_{3})\over -
\mu^{4}-\mu^{3} \lambda g_{3}+\mu^{2} \lambda^{2} g_{2}-\mu
\lambda^{3} g_{1}+\lambda^{4}g_{0}}\,, \nonumber
\\
\lambda_{3} = {\lambda^{3}\over - \mu^{4}-\mu^{3} \lambda
g_{3}+\mu^{2} \lambda^{2} g_{2}-\mu \lambda^{3}
g_{1}+\lambda^{4}g_{0}}\,. \nonumber \label{9.6}
\end{eqnarray}

We introduce new notation
\begin{eqnarray}
g_{0} = p_{4} , \; g_{1} = p_{3}, \; g_{2} = p_{2}, \; g_{3} =
p_{1} \; , \nonumber
\\
G^{4} =  p_{1} G^{3}  +  p_{2} G^{2}   + p_{3} G  +  p_{4} \; ,
\label{9.7}
\end{eqnarray}

\noindent then
\begin{eqnarray}
\lambda_{0} = {\mu^{3}+\mu^{2} \lambda p_{1}-\mu \lambda^{2}
p_{2}+\lambda^{3} p_{3}\over \mu^{4} + \mu^{3} \lambda p_{1} -
\mu^{2} \lambda^{2} p_{2} + \mu \lambda^{3} p_{3} - \lambda^{4}
p_{4}}\,, \nonumber
\\
\lambda_{1} = {-\mu^{2} \lambda-\mu ,\lambda^{2} p_{1}+\lambda^{3}
p_{2}\over \mu^{4} + \mu^{3} \lambda p_{1} - \mu^{2} \lambda^{2}
p_{2} + \mu \lambda^{3} p_{3}- \lambda^{4} p_{4}}\,, \nonumber
\\
\lambda_{2} = {\mu \lambda^{2}+\lambda^{3} p_{1}\over \mu^{4} +
\mu^{3} \lambda p_{1} - \mu^{2} \lambda^{2} p_{2} + \mu
\lambda^{3} p_{3}- \lambda^{4} p_{4}}\,, \nonumber
\\
\lambda_{3} = {-\lambda^{3}\over \mu^{4} + \mu^{3} \lambda p_{1} -
\mu^{2} \lambda^{2} p_{2} + \mu \lambda^{3} p_{3} - \lambda^{4}
p_{4}} \,; \nonumber
\\
\label{9.8}
\end{eqnarray}

\noindent recall that
\begin{eqnarray}
(\Lambda^{-1})_{\rho}^{\;\;\alpha} = \lambda_{0} + \lambda_{1} G +
\lambda_{2} G^{2} + \lambda_{3} G^{3} \; . \nonumber
\end{eqnarray}

Degrees of the matrix $G$ we can associate the following
invariants (see Chap. IV in \cite{Gantmakher}):
\begin{eqnarray}
 \mbox{Sp} ( G  ) =  g_{1} +g_{2} + g_{3} + g_{4} = s_{1} \; ,\quad
  s_{1} = G_{\alpha} ^{\;\; \alpha}(x)  \; ,
\nonumber
\\
\mbox{Sp} ( G ^{2} ) =  g^{2}_{1} +g^{2}_{2} + g^{2}_{3} +
g^{2}_{4} = s_{2} \;, \quad
 s_{2} = G_{\alpha}^{\;\;\rho}(x)
G_{\rho}^{\;\; \alpha} (x) \; , \nonumber
\\
 \mbox{Sp} ( G ^{3} ) =  g^{3}_{1} +g^{3}_{2} + g^{3}_{3} + g^{3}_{4} =s_{3} \; ,
 \quad
  s_{3} = G_{\alpha}^{\;\;\rho}(x) G_{\rho}^{\;\; \sigma}(x)  G_{\sigma}^{\;\; \alpha} (x)\; ,
\nonumber
\\
\mbox{Sp} ( G^{4}  ) =  g^{4}_{1} +g^{4}_{2} + g^{4} _{3} +
g^{4}_{4} =s_{4}\; , \quad
  s_{4} = G_{\alpha}^{\;\;\rho}(x) G_{\rho}^{\;\; \delta}(x)  G_{\delta}^{\;\; \sigma}(x)  G_{\sigma}^{\;\; \alpha} (x);
\label{9.9}
\end{eqnarray}

\noindent from these place the quantities $g_{1}, ..., g_{4}$
stand for  four eigenvalues of the matrix $G$.

Invariants  $s_{i}$ and $p_{i}$  obey the  Newton   recurrence
formulas  \cite{Gantmakher}:
\begin{eqnarray}
p_{1} = s_{1} =  \mbox{Sp} ( G  ) \;, \quad
p_{2} = {1\over 2} (s_{2} - p_{1}s_{1})  = {1\over 2} \left [
\mbox{Sp} ( G ^{2} )- p_{1} \mbox{Sp} ( G  ) \right ] \;,
\nonumber
\\
p_{3} = {1 \over 3} (s_{3} - p_{1}s_{2} -p_{2} s_{1} )
=  {1 \over 3} \left [ \mbox{Sp} ( G^{3}  ) - p_{1} \mbox{Sp} (
G^{2}  ) -p_{2} \mbox{Sp} ( G  ) \right ] \;, \nonumber
\\
p_{4} = {1 \over 4} ( s_{4} -p_{1} s_{3} - p_{2} s_{2} - p_{3}
s_{1})
= {1 \over 4}  [  \mbox{Sp} ( G^{4}   )  -p_{1} \mbox{Sp} (
G^{3}   ) - p_{2} \mbox{Sp} ( G ^{2} )- p_{3} \mbox{Sp} ( G  )
] . \nonumber
\end{eqnarray}

\noindent From this it follow the following representations for
the invariants  $p_{i}$:
\begin{eqnarray}
p_{1} = \mbox{Sp} ( G  ) \;, \quad
p_{2} =  {1\over 2}  \mbox{Sp} ( G ^{2} )- {1 \over 2}
\mbox{Sp}^{2} ( G  )  \;, \nonumber
\\
p_{3} =  {1 \over 3} \left [ \mbox{Sp} ( G^{3}  ) - \mbox{Sp} ( G
) \mbox{Sp} ( G^{2}  )  - {1\over 2} \left ( \mbox{Sp} ( G ^{2} )-
\mbox{Sp}^{2} ( G  ) \right )  \mbox{Sp} ( G  ) \right ] \nonumber
\\
= {1 \over 3}
 \mbox{Sp} ( G^{3}  )  -
{1\over 2} \mbox{Sp} ( G ^{2} ) \mbox{Sp} ( G  ) + {1 \over 6 }
\mbox{Sp}^{3} ( G  )    \;, \nonumber
\end{eqnarray}
\begin{eqnarray}
p_{4} =  {1 \over 4} \left [  \mbox{Sp} ( G^{4}   )  - \mbox{Sp} (
G  ) \mbox{Sp} ( G^{3}   )  - {1\over 2} \mbox{Sp}^{2} ( G ^{2} ) +{1 \over 2}
\mbox{Sp}^{2} ( G  )   \mbox{Sp} ( G ^{2} )   \right. \nonumber
\\
\left.  - {1 \over 3}
 \mbox{Sp} ( G^{3}  ) \mbox{Sp} ( G  ) +
{1\over 2} \mbox{Sp} ( G ^{2} ) \mbox{Sp}^{2} ( G  )
  -
 {1 \over 6 }  \mbox{Sp}^{3} ( G  ) \mbox{Sp} ( G  )
    \right ]  ;
\nonumber
\end{eqnarray}

\noindent finally for $p_{4}$ we find the expression
\begin{eqnarray}
p_{4} = {1 \over 4} \left  [  \mbox{Sp} ( G^{4}   )  - {4 \over 3}
\mbox{Sp} ( G  ) \mbox{Sp} ( G^{3}   )
 - {1\over 2} \mbox{Sp}^{2}
( G ^{2} ) +  \mbox{Sp}^{2} ( G  )   \mbox{Sp} ( G ^{2} )  -
 {1 \over 6 }  \mbox{Sp}^{4} ( G  )
   \right ] \; .
\nonumber
\end{eqnarray}

In the case when the matrix $ G $ is antisymmetric, we have the
equality
\begin{eqnarray}
\tilde{G} = - G, \; p_{1} = \mbox{Sp} G=0, \nonumber
\\
 \tilde{G^{3}
}= - G^{3}, \qquad  \mbox{Sp} ( G^{3} ) =0\;, \nonumber
\\
p_{1}=0,\qquad  p_{2} = {1 \over 2  } \mbox{Sp} (G^{2}), \nonumber
\\
 p_{3}=0,
 \;
 p_{4} = {1\over 4} \mbox{Sp} (G^{4}) + {1 \over 8} \mbox{Sp} ^{2}( G^{2}) \;.
 \label{9.12a}
 \end{eqnarray}

\noindent and the characteristic equation (9.7) takes the form
\begin{eqnarray}
G^{4} -   p_{2} G^{2}    -  p_{4} = 0 \;, \label{9.12b}
\end{eqnarray}
\noindent this case is realized in the construction of the
characteristic polynomial for the electromagnetic tensor. In this
case, the relations (\ref{9.8}) become more simple
\begin{eqnarray}
\lambda_{0} = {\mu^{3} -\mu\,\lambda^{2}\,p_{2}\over \mu^{4}  -
\mu^{2}\,\lambda^{2}\,p_{2}  - \lambda^{4}\,p_{4}}\,, \nonumber
\\
\lambda_{1} = {-\mu^{2}\,\lambda +\lambda^{3}\,p_{2}\over \mu^{4}
- \mu^{2}\,\lambda^{2}\,p_{2} - \lambda^{4}\,p_{4}}\,, \nonumber
\\
\lambda_{2} = {\mu\,\lambda^{2}\over \mu^{4}  -
\mu^{2}\,\lambda^{2}\,p_{2} - \lambda^{4}\,p_{4}}\,, \nonumber
\\
\lambda_{3} = {-\lambda^{3}\over \mu^{4}  -
\mu^{2}\,\lambda^{2}\,p_{2}  - \lambda^{4}\,p_{4}}\,.
\label{9.12c}
\end{eqnarray}

For additional verification we consider the simple case  in
absence of electromagnetic tensor when  additionally the
space-time is described by the Ricci tensor  of the following
simple form (elementary examples are the de Sitter spaces)
\begin{eqnarray}
G_{\alpha \beta} = {R \over 4} g_{\alpha \beta}  , \;
G_{\alpha}^{\;\; \beta} =
 {R \over 4} \delta_{\alpha}^{\;\; \beta},
\nonumber
\\
\mbox{Sp} G = R, \qquad  \mbox{Sp} (G^{2} ) = {1 \over 4} R^{2},
\nonumber
\\
 \mbox{Sp} (G^{3} ) = {1 \over 4^{2}} R^{3} ,\qquad
\mbox{Sp} (G^{4} ) = {1 \over 4^{3}} R^{4} , \label{9.13a}
\end{eqnarray}

\noindent that is
\begin{eqnarray}
p_{1} = R \;, \quad
p_{2} =  {1\over 2}  {1 \over 4} R^{2}  - {1 \over 2} R^{2} =
-{3\over 8 }  R^{2} \;, \qquad\qquad \nonumber
\\
p_{3} =   {1 \over 3}
 {1 \over 16 }  R^{3}   -
{1\over 2} {1 \over 4} R^{2} R + {1 \over 6 }  R^{3}   =  {1 \over
16 }   R^{3}\;,  \qquad\qquad \nonumber
\\
p_{4} = {1 \over 4} \left [  {1 \over 16 \cdot 4 } R^{4}  - {4
\over 3} R  {1 \over 16 } R^{3} - {1\over 2} {1 \over 16} R^{4}
 + R^{2}  {1 \over 4} R^{2}  -
 {1 \over 6 }   R^{4}
   \right ]  =  - { 1 \over 4^{4}} R^{4} \; .
\label{9.13b}
\end{eqnarray}

\noindent Expressions for $p_{i}$ correspond to the following
characteristic equation
\begin{eqnarray}
G = (G_{\alpha}^{\;\;\beta}), \qquad ( G -{R\over 4} )^{4} = 0 \;.
\label{9.13b}
\end{eqnarray}

Note that in the case of the presence of a geometric background
\begin{eqnarray}
  D_{\beta} \Phi - \lambda  \left (    F_{\beta \alpha } + i {\hbar c \over e}
   {R \over 4} g_{\beta \alpha }  \right )
  \Phi^{\alpha}
  - \mu   \Phi_{\beta}=0\; ,
  \nonumber
  \\
 D^{\beta} \Phi_{\beta} - \mu  \Phi =0 \; ;\qquad \qquad
\label{9.14a}
\end{eqnarray}

\noindent and in the absence of an external electromagnetic field,
the system of equations (\ref{9.14a}) looks as follows
\begin{eqnarray}
 D^{\beta} \Phi_{\beta} = Mc  \Phi \; ,\qquad
  D_{\beta} \Phi  =  \left ( Mc  + i\lambda    {\hbar c \over e} {R \over 4}
        \right )\Phi_{\beta}\,.
\label{9.14b}
\end{eqnarray}

In particular, in the case of the de Sitter spaces ($ R (x) = R $)
is equivalent to an effective additive (with a plus or minus) to
the mass of the particle
\begin{eqnarray}
 D^{\beta} \Phi_{\beta} = Mc  \Phi \; , \quad
  D_{\beta} \Phi  =  \left ( Mc  + i\lambda    {\hbar c \over e} {R \over 4}
        \right )\Phi_{\beta}\; ;
\label{9.14c}
\end{eqnarray}

\section{Conclusion}

Relativistic theory of the Cox's  scalar  not point-like particle with
intrinsic  structure is developed on the background of arbitrary curved space-time.
It is shown that in the most general form,  the  extended  Proca-like tensor first order
 system of equations contains non minimal interaction terms through electromagnetic tensor $F_{\alpha \beta}$
 and Ricci tensor $R_{\alpha \beta}$.

 In
relativistic Cox's theory,  the limiting procedure to
non-relativistic approximation is performed in a special class of  curved
space-time models.
   This theory is specified in simple geometrical backgrounds:
   Euclid's, Lobache\-vsky's, and Rie\-mann's.
  Wave equation  for the Cox's particle
is solved  exactly in presence of external uniform magnetic and
electric fields in the case of Minkowski space. Non-trivial
additional  structure of the particle modifies the frequency of a
quantum oscillator arising effectively in presence if external
magnetic field. Extension of these problems to the case of the
hyperbolic Lobachevsky  space is examined.
 In
presence of  the magnetic field, the quantum problem in radial
variable has been solved exactly; the quantum motion in
z-direction is described by 1-dimensional Schr\"{o}dinger-like
equation in an effective potential which turns out to be  too
difficult for analytical treatment. In the presence of electric
field, the situation is similar. The same analysis has been
performed for spherical Riemann space model.

General conclusion
can be done:  the effects  of large scale structure of the Universe
depends greatly on the form of basic  equations for elementary
particle,
 any modifications of them lead to new physical phenomena due to non-Euclidean
 geometry background.

\section{Acknowledgements}

This  work was   supported   by the Fund for Basic Researches of
Belarus,
 F 13K-079, within the cooperation framework between Belarus  and Ukraine.
 Authors are grateful  to organizers of
 The  International conference
in honor of Ya. B. Zeldovich 100-th Anniversary:  Subatomic
particles, Nucleons, Atoms, Universe: Processes and Structure,
held  March 10--14, 2014, Minsk, Belarus for possibility  to give
this talk.

\end{document}